\documentclass[a4paper,usenatbib]{mn2e}
\usepackage{subfloat}
\usepackage{times}
\usepackage{natbib}
\usepackage[tableposition=top]{caption}
\makeatletter
\renewcommand\hyper@natlinkbreak[2]{#1}
\makeatother
\usepackage{natbib,hyperref,ifthen} 
\usepackage{hyperref}
\usepackage{etoolbox}
\hypersetup{
 colorlinks=true,
 citecolor=blue}
 \usepackage[T1]{fontenc}
 \usepackage{aecompl}
\usepackage{caption}
\usepackage{subfigure}
\usepackage{sidecap}
 \usepackage{graphicx}
 \usepackage{graphics}
 \usepackage{amsmath}
  \usepackage{lipsum} 
 \usepackage{afterpage}
 
 \newcommand{\aap}{A\&A}

\newcommand{\aj}{AJ}
\newcommand{\apj}{ApJ}
\newcommand{\apjs}{ApJS}
\newcommand{\apjl}{ApJL}

\newcommand{\mnras}{MNRAS}
\newcommand{\nat}{nature}

 \renewcommand\appendix{\par
  \setcounter{section}{0}
  \setcounter{subsection}{0}
  \setcounter{figure}{0}
  \setcounter{table}{0}
  \renewcommand\thesection{ \Alph{section}}
  \renewcommand\thefigure{\Alph{section}\arabic{figure}}
  \renewcommand\thetable{\Alph{section}\arabic{table}}
}

\title[Non-circular flows in barred galaxies]{Estimating non-circular motions in barred galaxies using numerical N-body simulations}
\author[ T. Randriamampandry, F. Combes, C. Carignan \& N. Deg]{T. H. Randriamampandry$^{1}$\thanks{tokyr@ast.uct.ac.za}, F. Combes$^{2}$, C. Carignan$^{1}$ and N. Deg$^{1}$\\
$^{1}$ Department of Astronomy, University of Cape Town, Private Bag X3, Rondebosch 7701, South Africa\\ 
$^{2}$Observatoire de Paris, LERMA, CNRS, 61 Av. de l'Observatoire, F-75014 Paris, France}

\begin{document}

\date{}

\pagerange{\pageref{firstpage}--\pageref{lastpage}} \pubyear{2015}

\maketitle
\label{firstpage}

\begin{abstract}
The observed velocities of the gas in barred galaxies are a combination of the azimuthally-averaged circular velocity and non-circular motions, primarily caused by gas streaming along the bar. These non-circular flows must be accounted for before the 
observed velocities can be used in mass modeling.  In this work, we examine the performance of the tilted-ring method and the DiskFit algorithm 
for transforming velocity maps of barred spiral galaxies into rotation curves (RCs) using simulated data.  We find that the tilted-ring 
method, which does not account for streaming motions, under/over-estimates the circular motions when the  bar is parallel/perpendicular to the projected major axis.  DiskFit, which does include streaming motions, is limited 
to orientations where the bar is not-aligned with either the major or minor axis of the image.  Therefore, we propose a method of correcting RCs based on numerical simulations of galaxies.  We correct the RC derived from the tilted-ring method based on a numerical simulation of a galaxy with similar properties and projections as the observed galaxy.  Using observations of NGC 3319, which has a bar aligned with the major axis, as a test case, we show that 
the inferred mass models from the uncorrected and corrected RCs are significantly different.  These results show the 
importance of correcting for the non-circular motions and demonstrate that new methods of accounting for these motions are necessary as current 
methods fail for specific bar alignments.

%
. 
\end{abstract}


\section{Introduction}

 The rotational velocities obtained from gas observations are one of the most 
commonly used tools to study the mass distributions of spiral galaxies.  It is often assumed that the gaseous motions are purely circular
due to their low velocity dispersions.  However, the presence of a galactic bar will induce non-circular motions due to the streaming along the bar. Since bars account for two-thirds of nearby galaxies (eg. \citealt{1991rc3..book.....D, 2000AJ....119..536E, 2000ApJ...529...93K, 2002MNRAS.336.1281W, 2007ApJ...659.1176M, 
2007ApJ...657..790M, 2008ASPC..396..351B, 2009A&A...495..491A, 2009ASPC..419..138M, 2010ApJ...711L..61M, 2011MNRAS.411.2026M}), it is important to properly account for these motions.


The study of the mass distribution is done by comparing the observed RC
 with the expected contributions from the gas, stars and dark matter,
assuming axisymmetry, and that the gas is in circular motions (e.g. \citealt{de-Blok:2008oq}). For the dark matter component,  an empirical or 
theoretical density profile is used. There are several dark matter density profiles suggested in the literature. The two most commonly used are the 
NFW profile \citep{nfw96} derived from cosmological simulations with a central cusp and the isothermal (ISO) model
(see e.g: \citealt{1985ApJ...299...59C, 1985ApJ...294..494C}) with a central flat core (see also: \citealt{1969AN....291...97E, 1995ApJ...447L..25B, 1998ApJ...502...48K}), a more general form of the dark matter density profile is also given by \cite{2001AJ....121.1952B}. 
 Alternative model such as the Modified Newtonian Dynamics (MOND) \citep{Milgrom:1983lq, Milgrom:1983rr, Milgrom:1983rd} do not have a dark matter component. \cite{Milgrom:1983rr} stipulated that there is no need for a dark matter halo to explain the flatness of galaxies' rotation curves (RCs) if the law of gravity is modified below a critical acceleration a$_{0}$.
 A comparison between ISO dark matter models and MOND models, for a sample of dwarf and spiral galaxies, is given in \cite{2014MNRAS.439.2132R} but a larger sample is needed before a final conclusion can be made.

  In order to distinguish between both these and other models, RCs that represent the gravitational potential are 
necessary.  RCs are often derived from observed velocity maps of gas using the tilted ring method \citep{1974ApJ...193..309R}.
This method divides the velocity map into co-centered rings and infers a variety of kinematic parameters, including the circular 
velocities.  It has been implemented in  the GIPSY \citep{2011ascl.soft09018A} task {\sc rotcur} \citep{1989A&A...223...47B}.
Unfortunately, { \sc rotcur} does not correct for the non-circular motions.  There are other methods that have been developed to account 
for these motions, including the publicly available DiskFit code \citep{2007ApJ...664..204S}. DiskFit creates physical models from the observed rotation velocities and extract the non-circular motions 
using a $\chi^{2}$ minimization technique (for more details see : \footnote{\url{http://www.physics.rutgers.edu/~spekkens/diskfit/}}). DiskFit has proven to be able to detect weak bars \citep{2012MNRAS.427.2523K}. It was specifically designed to fit a non-axisymmetric flow pattern to 
two-dimensional velocity maps \citep{2007ApJ...664..204S}. However, it is well known that the DiskFit bar flow algorithm fails when the bar is almost parallel to the major or minor axis because of the degeneracy of the velocity components \citep{2010MNRAS.404.1733S}.

 \cite{2008MNRAS.385..553D} used numerical models with pure rotation and compared the known circular velocities with the observed rotational velocities. They found that if a bar is perpendicular to the major axis of a galaxy, 
the measured rotational velocities and thus the mass is over estimated and the contrary if the bar is parallel to the major axis.%
%

  We propose an alternate method to account for non-circular motions using N-body simulations.  A grid of models with different bar orientations, strengths, and bulge-disk-halo compositions is generated.  
RCs from images of these models are made using {\sc rotcur}, which are then compared to the RCs calculated from the gravitational potential and its derivatives.  This comparison gives the corrections necessary for non-circular motions. To get the corrected RC for a real galaxy, the best corresponding model 
from the grid of simulations is selected and the corrections derived from the simulation are applied to the actual galaxy's RC derived from {\sc rotcur}.

The paper is organised as follow:  Section \ref{data} describes the details of both the numerical simulations and the {\sc rotcur} and DiskFit algorithms.
Section \ref{res} compares the RCs derived from the simulations using {\sc rotcur} and DiskFit to the RC obtained from the 
gravitational potential.  In Section \ref{case} we describe our new method of correcting RCs and use NGC 3319 as a test 
case to show the importance of properly accounting for non-circular motions in mass models.  Finally, in Section 5 we present our conclusions.

\section{Data analysis}
\label{data}
\subsection{N-body simulations}
\label{sec:nbody}

We use a series of spiral galaxy simulations, with different Hubble types,
Sa, Sb, Sd, fitted to observed rotation velocities (cf \citealt{2010A&A...518A..61C}) and different radial distributions of dark matter, with either a core or a cusp
in the center, and different dark-to-visible mass ratios (maximum disk or not).
 They give a range of rotation velocities with various shapes, rising more
or less slowly with radius. We do not vary the total mass of the galaxies, since only the shapes of the rotation velocities are important, and masses can be varied by a change of units. It is acknowledged that there is a mass and size evolution along the Hubble sequence (see e.g. \citealt{2014ApJ...788...28V}), and that late-type spiral galaxies are lighter and smaller, but these can be taken into account afterwards through rescaling.
For all galaxy models, the total mass inside 35 kpc
is 2.4 10$^{11}$ M$_\odot$, with 75\% in dark matter and 25\% in baryons,
for the standard model (based on the $\Lambda$CDM cosmology) and equality between baryons and dark matter
for the maximum disk model.

\begin{table*}
\caption{Initial conditions for the various components (bulge, disk, halo, and gas) of the different models used. All models are normalized to giant galaxies to explore the different shapes. Dwarfs can be obtained through rescaling (see text).}
\label{tab:simul}
\begin{center}
\begin{tabular}{lcccccccc}
\hline
\\
Model & M$_b$ & r$_b$ & M$_d$ & r$_d$ & M$_h$ & r$_h$ & M$_{gas}$ & r$_{gas}$ \\
&10$^{10}$ M$_\odot$& kpc & 10$^{10}$ M$_\odot$  & kpc &  10$^{10}$ M$_\odot$&kpc &  10$^{10}$ M$_\odot$& kpc \\
\hline
\\
gSa & 4.6  & 2. &  6.9 & 4. &  11.5 & 10.  & 1.3  & 5. \\
gSb-st & 1.1  & 1. &  4.6 & 5. &  17.2 & 12.  & 0.9  & 6. \\
gSb-dmx & 2.2  & 1. &  9.2 & 5. &  11.4 & 12.  & 0.9  & 6. \\
gSb-nfw & 1.1  & 1. &  4.6 & 5. &  17.2 & 12.  & 0.9  & 6. \\
gSd-st & 0.  & -- &  5.7 & 6. &  17.2 & 15.  & 1.7  & 7. \\
gSd-dmx & 0.  & -- &  11.4 & 6. &  11.4 & 15.  & 1.7  & 7. \\
gSd-nfw & 0.  & -- &  5.7 & 6. &  17.2 & 15.  & 1.7  & 7. \\
\\
\hline
\end{tabular}
\end{center}
\end{table*}

 We have used two different simulation codes, in order to test
that our assumptions are not biasing the results. In both kinds
of simulations, the runs are fully self-consistent, with live dark haloes.
The first code is based on a 3D particle-mesh and the Fast-Fourier Transform  (FFT) technique to solve the Poisson equation, and the gas is represented by sticky particles (FFT-sticky particles). 
  We also ran a set of simulations using the Tree-SPH code used in the Galmer project \citep{2007A&A...468...61D}.  More details on these simulations including the hydrodynamics, star formation, and feedback can be found in Appendix A. The total number of particles is 240k in the FFT-sticky code and 480k in the Tree-SPH code. 

These produced different outputs for the gas particle distributions, but the effects of the bar on the observed velocity maps were the same.  Since the conclusions are the same, we will primarily examine the results for the FFT-stickly particles simulations and the results for the Tree-SPH runs can be found in Appendix B. 

%
%
%
%
%

%

 For the initial conditions of the galaxies of different Hubble types,
we have followed the procedure described by \cite{2007A&A...468...61D}.
The halo and the optional bulge are modeled as a Plummer sphere,
and the stellar and gaseous disks follow a Miyamoto-Nagai density
profile.

\begin{figure*}
\centerline{
\includegraphics[angle=-90,width=15cm]{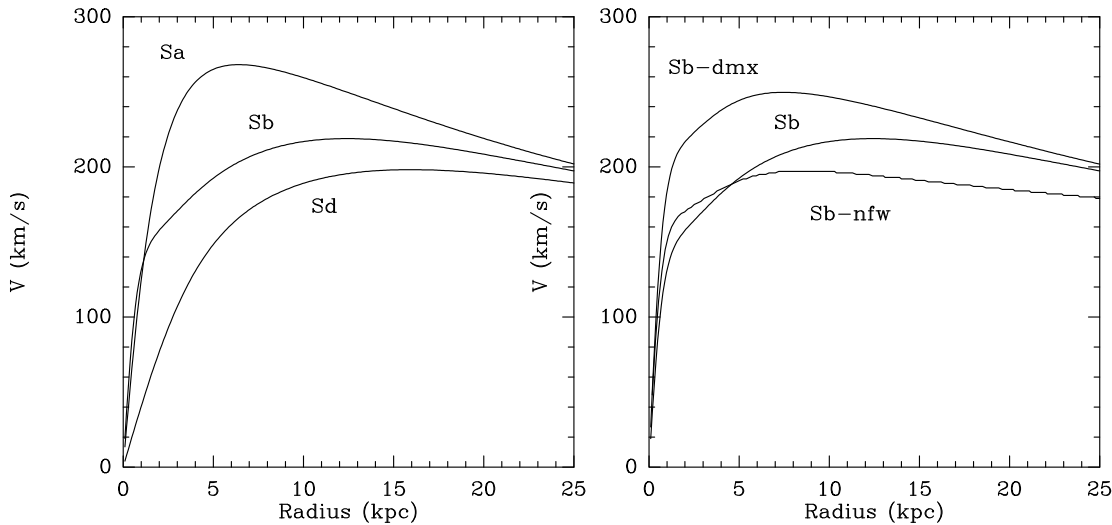}}
\caption{Initial RC
s of the various models, described
in table \ref{tab:simul}. The left plot compares the various morphological types, and the right plot the various dark matter distributions.
}
\label{fig:vrotall}
\end{figure*}

The mass and characteristic radii of all components in
the seven initial conditions, are displayed in Table \ref{tab:simul}.
The standard model (-st) has a disk dominated by dark matter,
with a central core. A NFW model, with a cusp and a central
concentration of c = 8  has been run (-nfw), then a maximum disk model
(-dmx) with equal mass between baryons and dark matter within the optical
disk. 
The initial rotational velocities, corresponding to the main variations
of the parameters, are displayed in figure \ref{fig:vrotall}.

The expected velocities  ($ \rm V_{expected}$) were calculated from the gravitational potential and its derivatives. More specifically, the potential derived from the disk
is averaged out azimuthally, to obtain $\Phi$(r), and the circular velocity $V_{circ}$ is

\begin{equation}
\rm V_{circ}^2 = - r \frac{d\left<\Phi\right> }{dr}
\end{equation}

\subsection{Snapshots}

The snapshots from the FFT-sticky particles simulation were extracted from the data using a projection code specifying the inclination angle and the position angle and deriving the radial velocity of the gas particles at each point of a data cube. The data cube have three axes, the first and second axis are the right ascension and declination and the third axis is the velocity in km/s. Each cubes have 50 channels with a 12.5 km/s velocity resolution.
The cube is then transformed into a fits file. The projection angle PA was varied between 10 to 180 degrees in order to cover a wide range of intrinsic bar orientations for all the snapshots.

  We select the snapshots where the bar has reached a quasi-stationary state (most of the gas have been converted into stars),
but is still strong enough, and there is enough gas remaining in the disk
to sample the rotation velocities with gas only. Indeed, the simulations do not take
into account gas accretion form the cosmic filaments, and the gas density is progressively
reduced by star formation.

\subsection{Measuring the rotational velocities}

\subsubsection{Using the tilted-ring method with {\sc rotcur}}

{\sc rotcur} uses the tilted-ring method \citep{1974ApJ...193..309R}, which divides  the observed velocity field into concentric rings and estimates the circular velocities and other kinematical parameters such as position angle (PA) and inclination using a $\chi^{2}$ minimization technique. This method assumes that the gas is moving purely along circular orbits. Each ring is characterized by the rotation center x$_{\rm c}$  and y$_{\rm c}$, the systemic 
velocity V$_{\rm sys}$, the expansion velocity V$_{\rm exp}$ (assumed here to be zero), the inclination angle and PA. These parameters can be fixed or let free to vary depending on the user.
 The line of sight velocity at any position in the ring is then given by:
 \begin{equation}
\rm V(x,y) = V_{sys} + V_{C}sin(i)cos(\theta),
\end{equation}
where $\theta$ is the position angle with respect to the receding major axis measured in the plane of the sky, V$_{\rm sys}$ the systemic velocity and i the inclination angle.

$\theta$ is related to the actual PA by the following equations:
\begin{equation}
\rm cos(\theta) = \frac{- (x - x_{c})sin(PA) + (y-y_{c})cos(PA)}{R}
\end{equation}
\begin{equation}
\rm sin(\theta) = \frac{- (x - x_{c})cos(PA) + (y-y_{c})cos(PA)}{Rcos(i)}
\end{equation}

The data cubes extracted from the simulation snapshots were analyzed with the Groningen Image Processing System (GIPSY: \url{http://www.astro.rug.nl/~gipsy/}) software. The task {\sc rfits} was used to convert the fits file into a GIPSY format, then the fits headers were changed using {\sc fixhed}. The following parameters were modified in the header; the axis and the pixel size which is necessary before extracting the velocity fields. The task {\sc momnt} was used to produce the moment maps, and the rotation velocities were derived using {\sc rotcur}. The expansion velocity is fixed to zero for all the models. We choose an inclination angle of 60 degrees which is suitable for the kinematical studies. The systemic velocity ( V$_{\rm sys}$) and position angle (PA) are fixed using the following values: V$_{\rm sys}$ = 0.0 km s$^{-1}$  and PA = 0.0 degree during the fitting procedures.

The process of deriving the RCs includes three steps; firstly, the PA and inclination are held fixed using the optical values and the systemic velocity and position centre (Xpos and Ypos) are allowed to vary; secondly, the systemic velocity and the kinematic centre are held fixed and the PA and inclination are let free to vary; finally, all the above parameters are fixed and the average values are used to derive the RCs.

\subsubsection{Using the DiskFit bisymmetric model}

 DiskFit \citep{2007ApJ...664..204S} is specifically designed to correct the non-circular flows due to bars. It creates physical models from the observed velocity fields and extracts sthe velocities and other parameters from the model velocity map. DiskFit also uses a bootstrap method to estimate the uncertainty on the parameters and uses all to data during the fitting procedures.

DiskFit compares the observed velocity field with the following equation:
\begin{equation}
\begin{split}
\rm V_{model} = V_{sys} + sin(i)[V_{t}cos(\theta) - V_{2,t}cos(2\theta_{b}) cos(\theta) \\\\
\rm - V_{2,r}sin(2\theta_{b}) sin(\theta)
\end{split}
\end{equation}
where V$_{t}$ is the circular velocity, V$_{2,t}$ and V$_{2,r}$ the amplitudes of the tangential and radial component of the noncircular motions for a bisymmetric flow model.
This form of the equation is for bisymmetric flows from bars.  Moreover, DiskFit can also deal with warped galaxies.  Regardless,
if the gas is moving on purely circular orbits, this equation reduces to the key equation of the tilted-ring method.

DiskFit estimates the rotational velocities and the other parameters using a $\chi^{2}$ minimization technique:

\begin{equation}
\rm \chi^{2} = \sum^{N}_{n=1}(\frac{V_{obs}(x,y)-\sum^{k}_{k=1}\omega_{k,n}V_{k}}{\sigma_{n}})^{2}
\end{equation}
where V$_{\rm obs}$(x,y) is the observed velocity at the position (x,y) on the sky, $\sigma_{n}$ is the uncertainty, $\omega_{k,n}$ is a weighting function which includes the trigonometric factors and also defines an interpolation scheme for the projected model \citep{2010MNRAS.404.1733S}. Some of the advantages of DiskFit are listed in \cite{2010MNRAS.404.1733S}. DiskFit uses all the data in a single fit, which is necessary to detect mild distortion and also gives a better estimate of the uncertainty using a bootstrap method \citep{2007ApJ...664..204S}. 
\section{Results and discussions}
\label{res}
\subsection{Results using {\sc rotcur}}

\begin{figure*}
\centerline{
\includegraphics[angle=0,width=22cm]{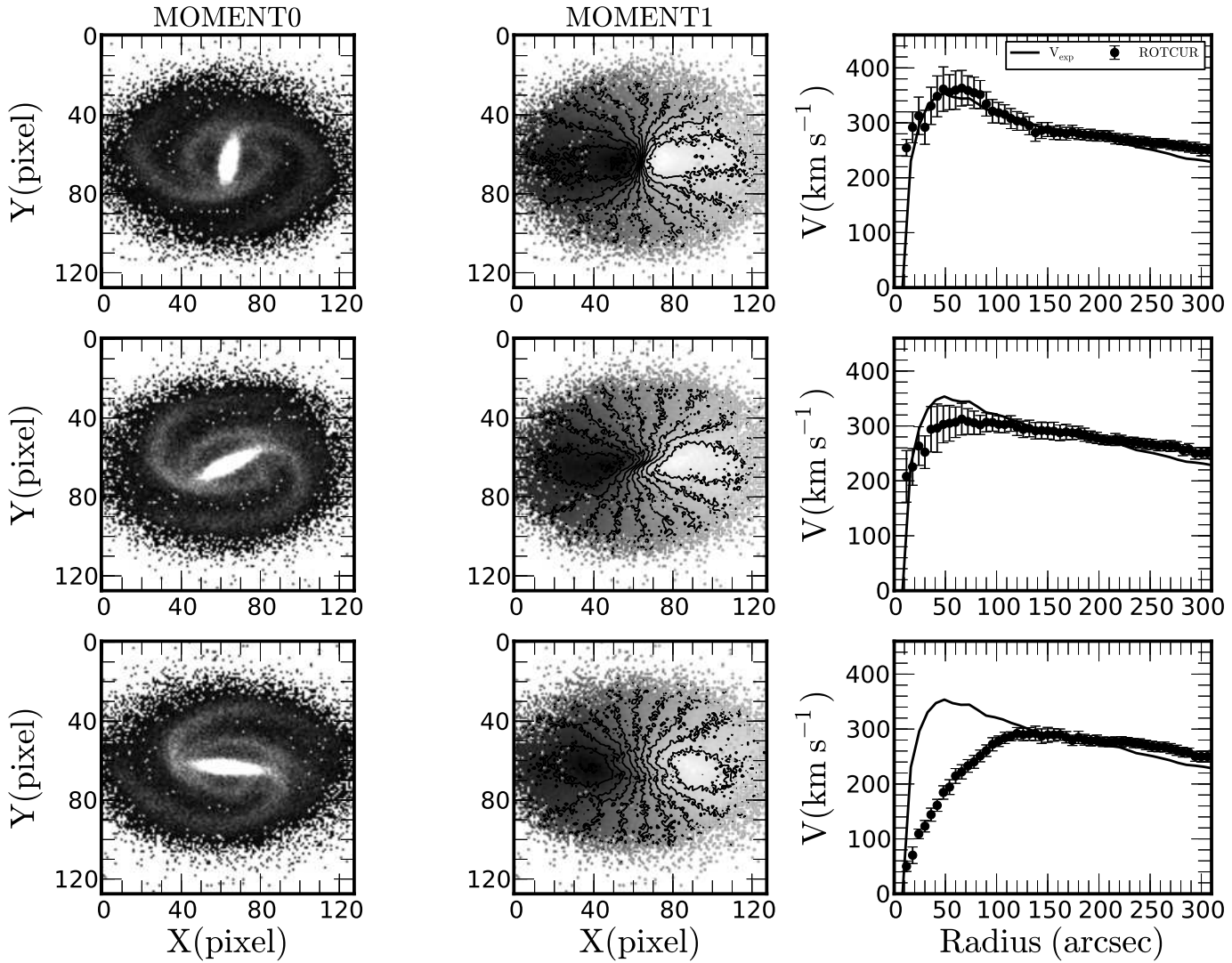}}
\caption{Three different bar positions for the model gSa (FFT-Sticky particles, T= 700 Myrs) 
Top panel: the bar is perpendicular to the major axis, middle panel: intermediate position and bottom panel: parallel to the major axis. 
The first column is the moment0 map, second column the moment 1 map superposed with the iso-velocity contours and the third 
column the comparison between the expected velocities shown as a continuous line and the measured rotation velocities derived using {\sc rotcur} (black points). The axes of the moment maps are in pixels  (1 pixel = 5 arcsec).}
\label{stck1}
\end{figure*}

\begin{figure*}
\centerline{
\includegraphics[angle=0,width=22cm]{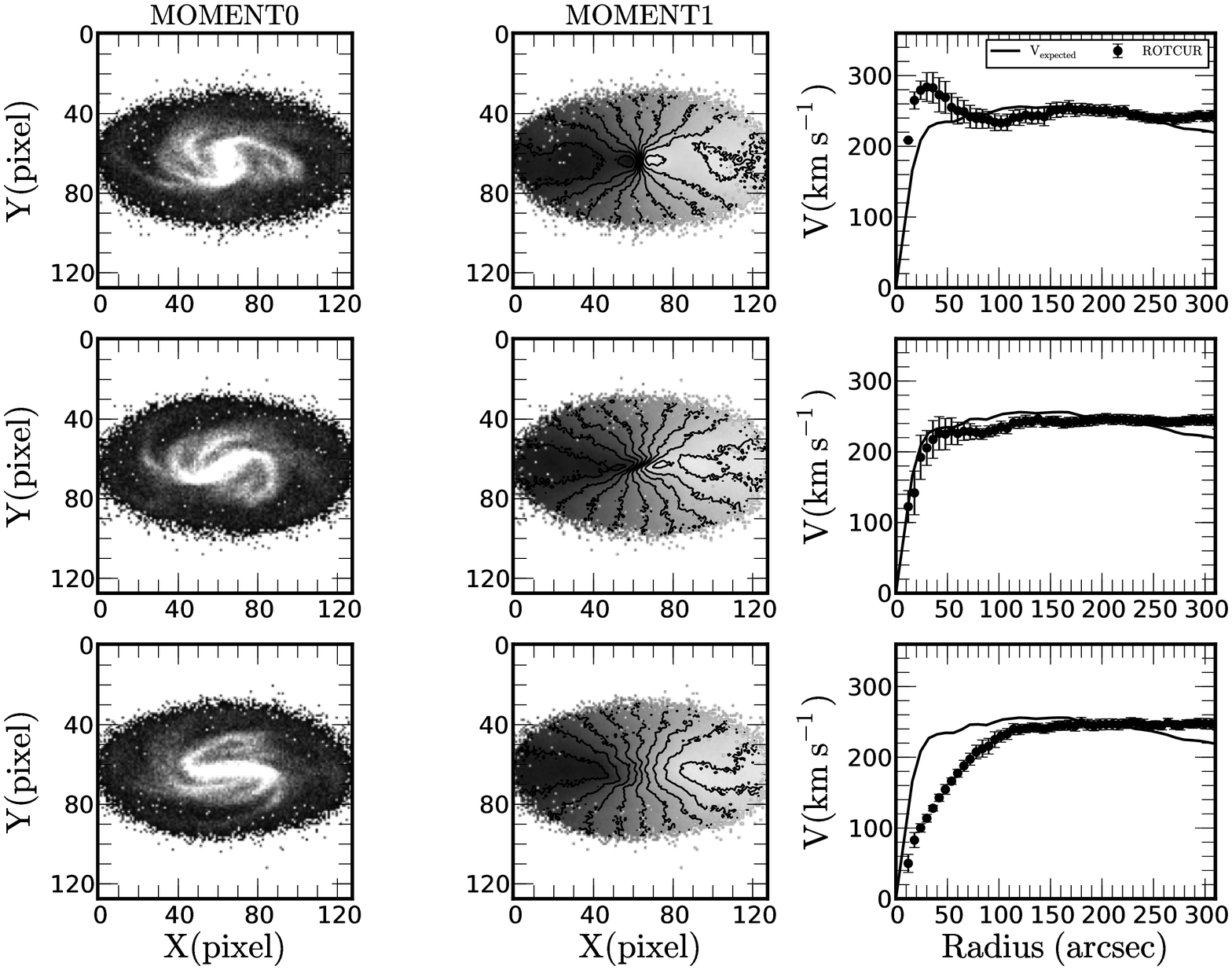}}
\caption{Three different bar positions for the model gSb (FFT-sticky particles , T = 700 Myrs see figure \ref{stck1} for details).}
\label{stck2}
\end{figure*}

\begin{figure*}
\centerline{
\includegraphics[angle=0,width=22cm]{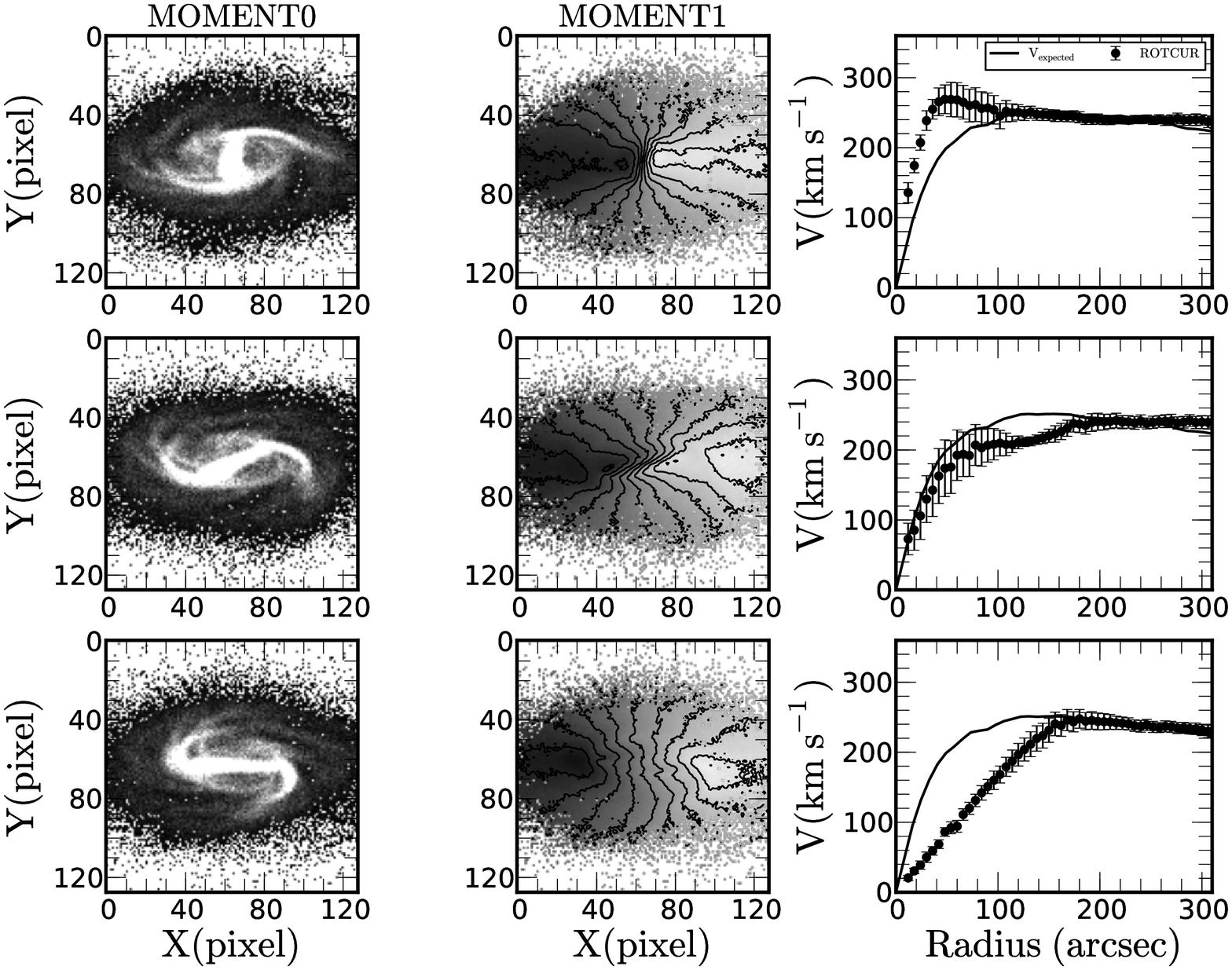}}
\caption{Three different bar positions for the model gSd ( FFT-sticky particles, T = 900 Myrs see figure \ref{stck1} for details).}
\label{stck3}
\end{figure*}

\begin{figure*}
\centerline{
\includegraphics[angle=0,width=16cm]{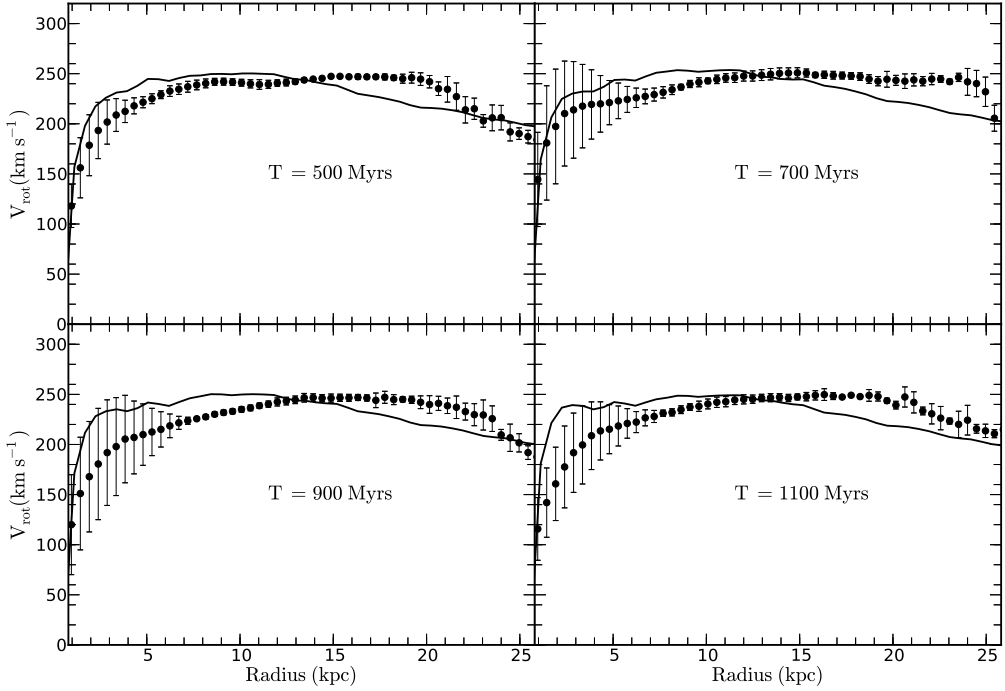}}
\caption{Standard gSb galaxy model: comparison between the expected RCs calculated from the gravitational potential as continuous lines and the average RCs for different bar orientations as filled circles with the 1-$\sigma$ error bars. The RCs of 12 simulated galaxies with different bar orientations were extracted and averaged for each snapshot (PA varies between 0 and 180 degrees with a step of 15 degrees). The epochs of the snapshots are shown on the graph.}
\label{fig:vrotstr}
\end{figure*}

 \begin{figure*}
\begin{minipage}{140mm}
  \begin{tabular}[t]{c}
         \hspace*{-6.0em} \vspace{-1.0em}
            \subfigure{\includegraphics[width=0.42\linewidth]{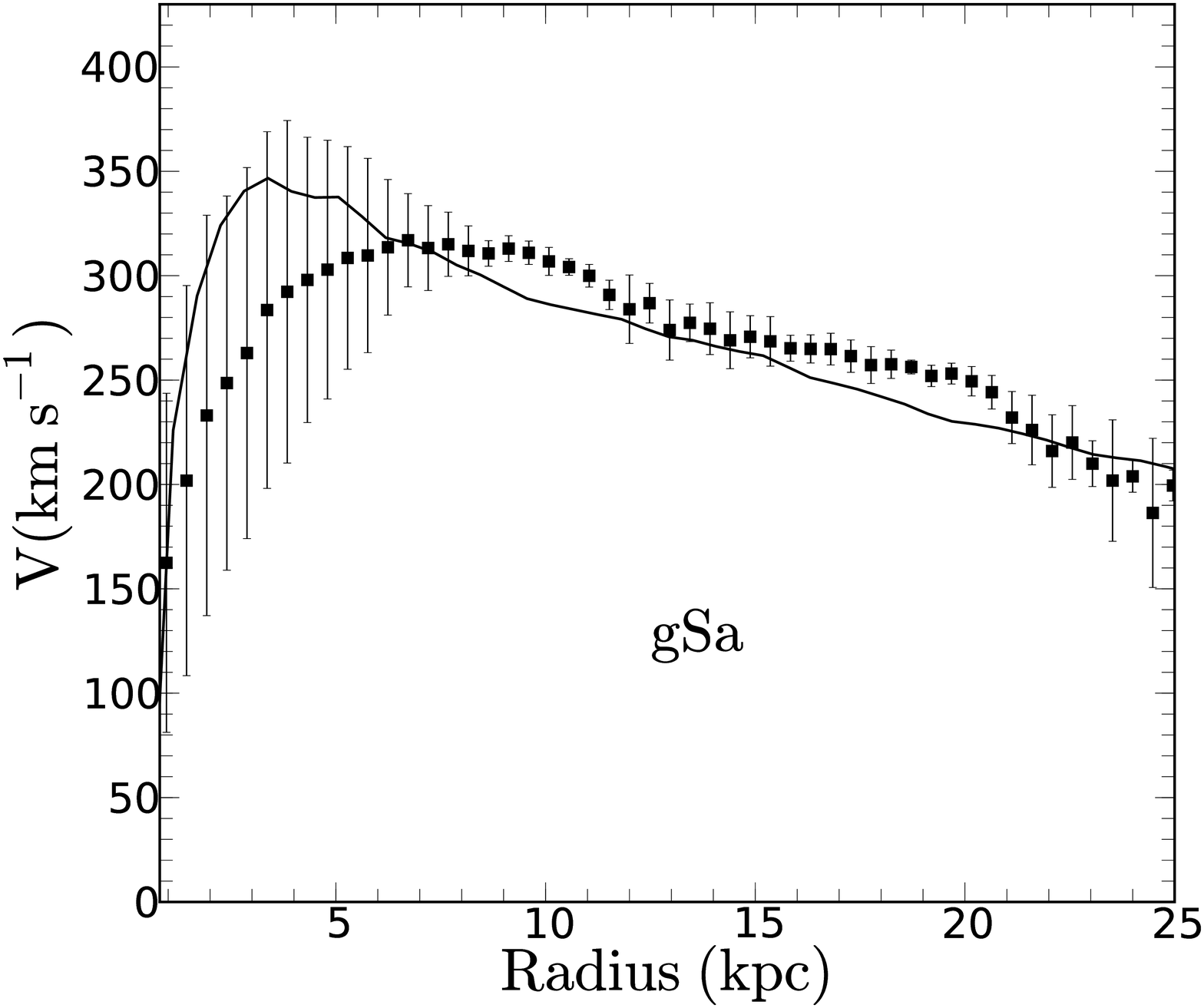}}\\
	   	\hspace*{-6.0em} \vspace{-1.0em}
        \subfigure{\includegraphics[width=0.42\linewidth]{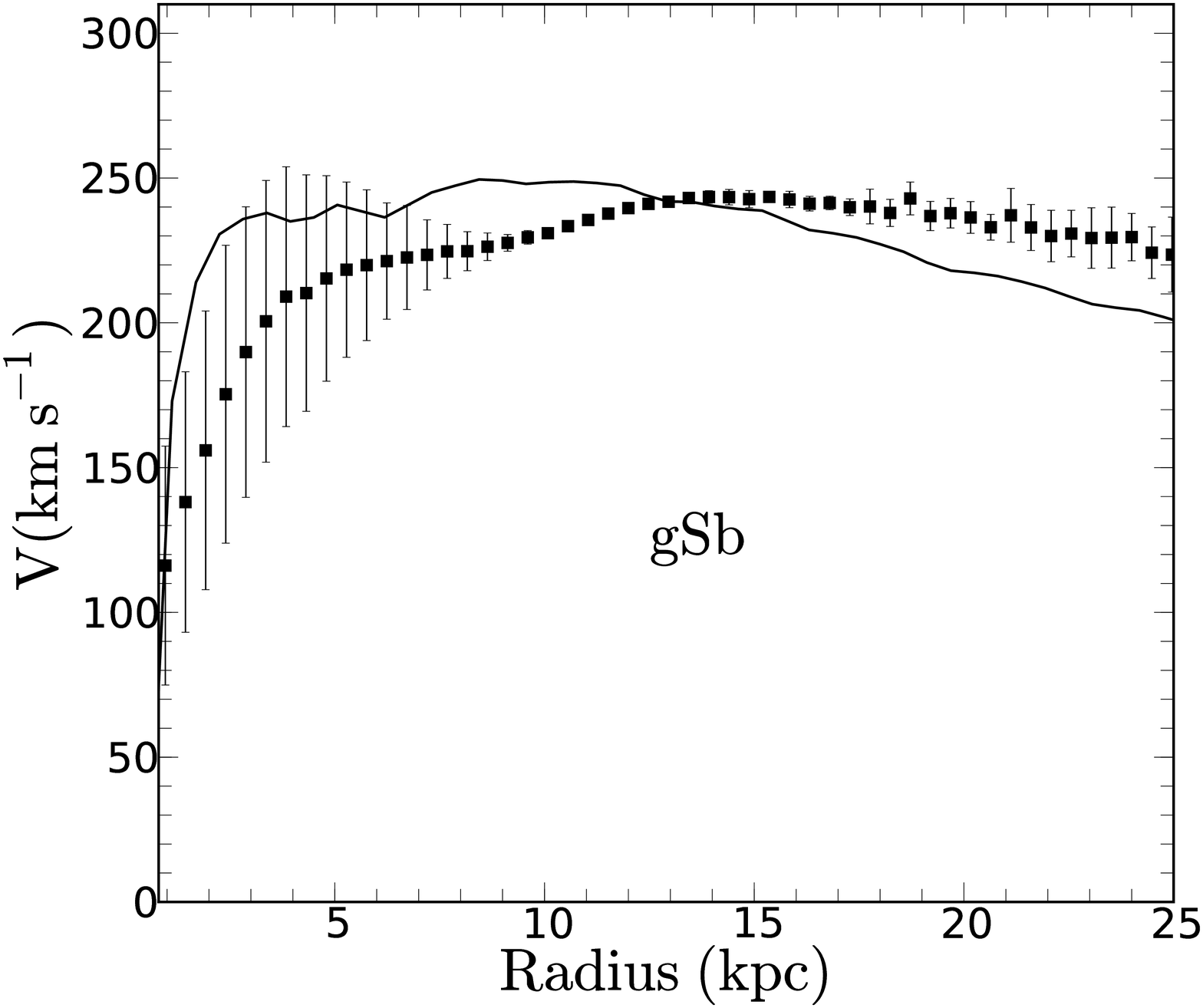}} 
         \hspace*{-1.0em}
        \subfigure{\includegraphics[width=0.42\linewidth]{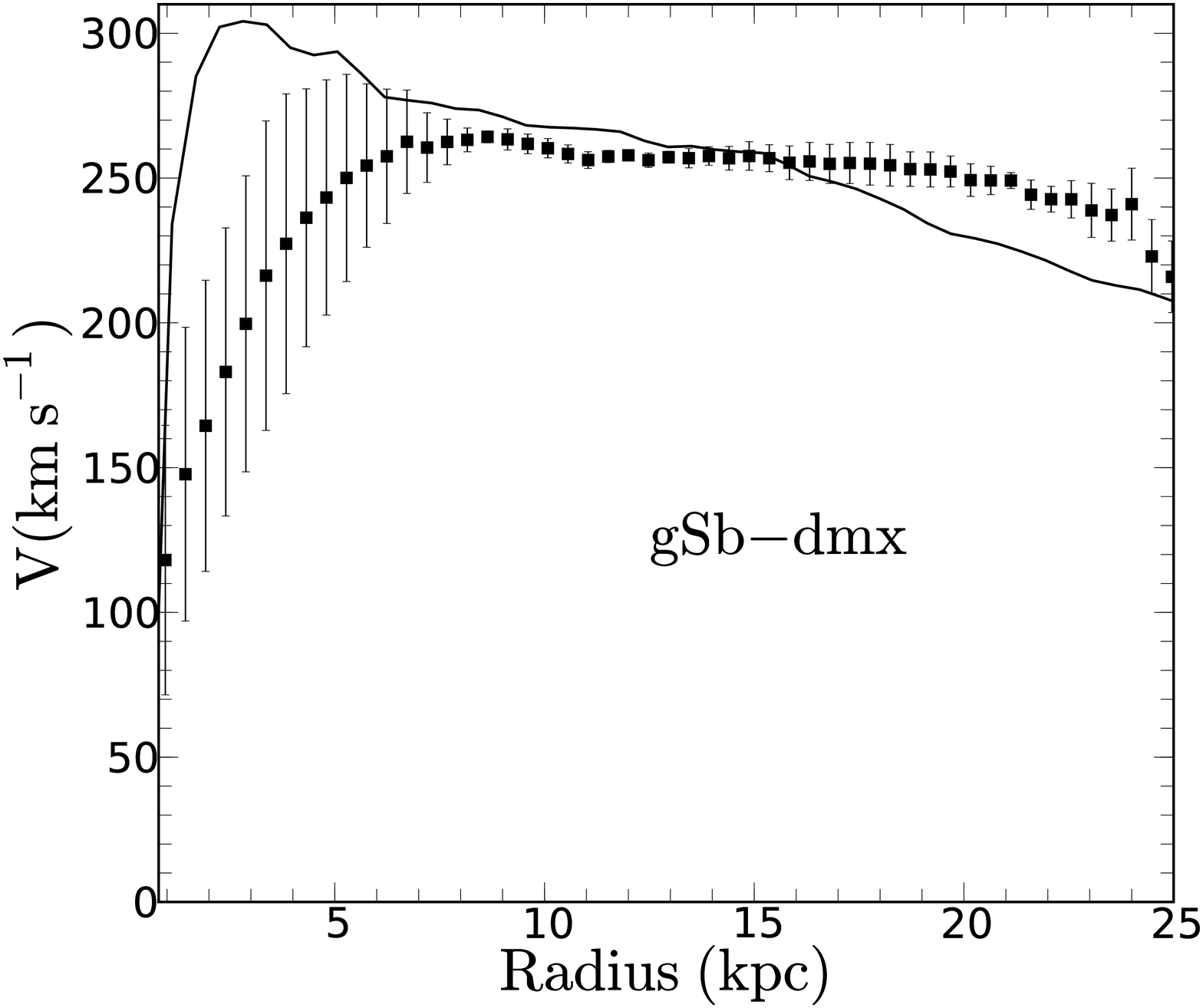}} 
        \hspace*{-1.0em} \vspace{-1.0em}
          \subfigure{\includegraphics[width=0.42\linewidth]{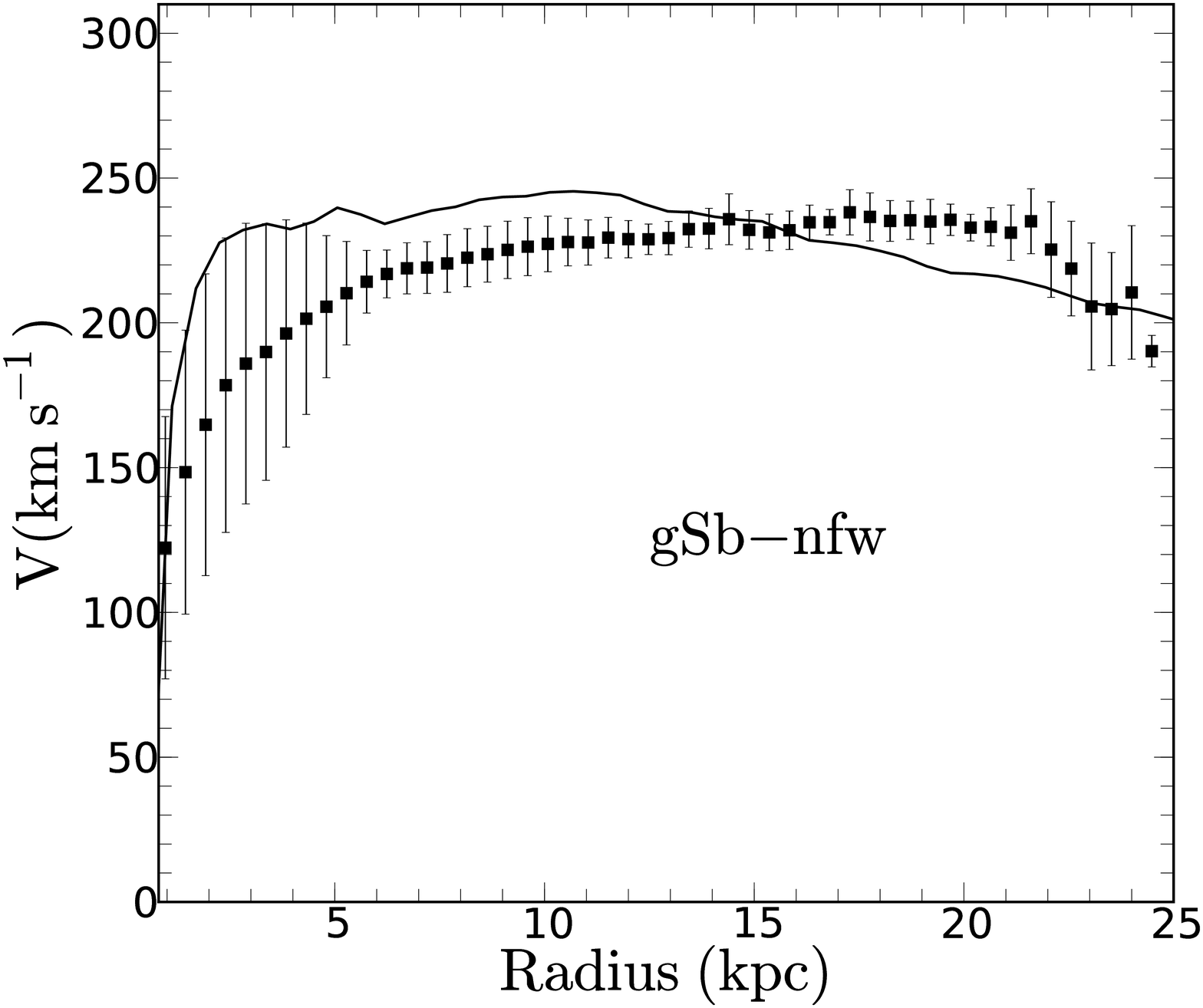}}\\
	   	\hspace*{-6.0em} \vspace{-1.0em}
        \subfigure{\includegraphics[width=0.42\linewidth]{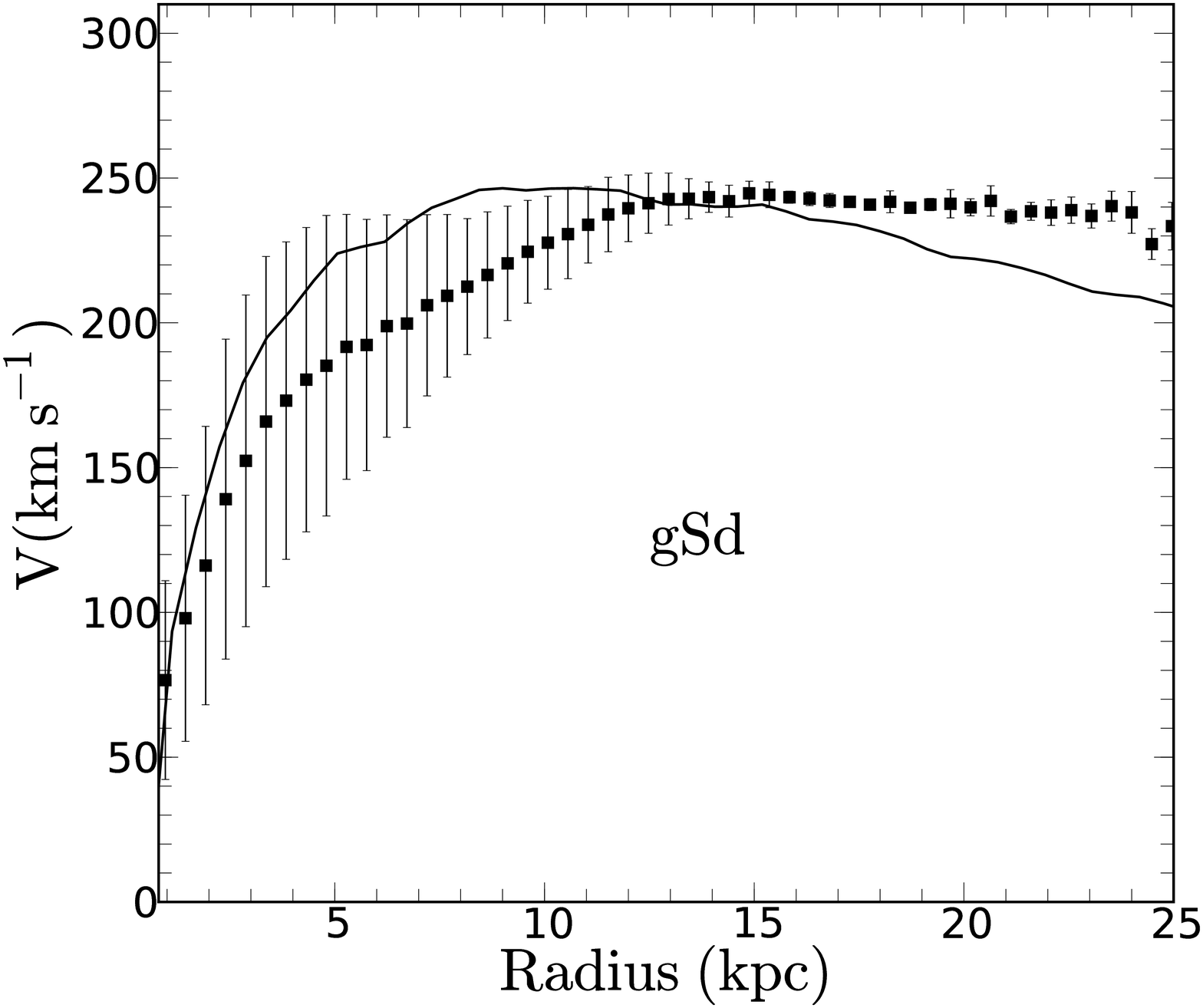}} 
         \hspace*{-1.em}
        \subfigure{\includegraphics[width=0.42\linewidth]{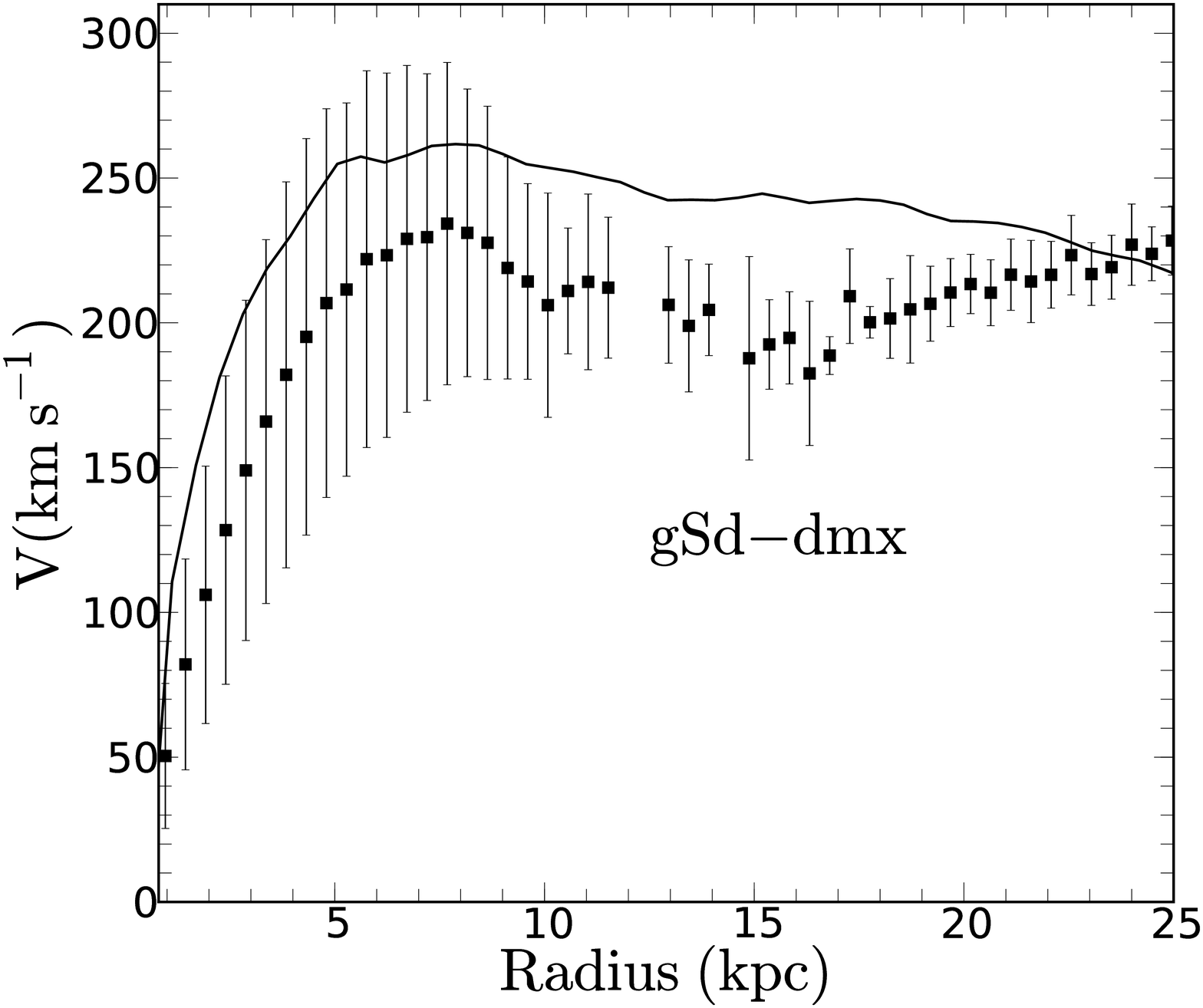}} 
        	\hspace*{-1.0em} \vspace{1.0em}
	  \subfigure{\includegraphics[width=0.42\linewidth]{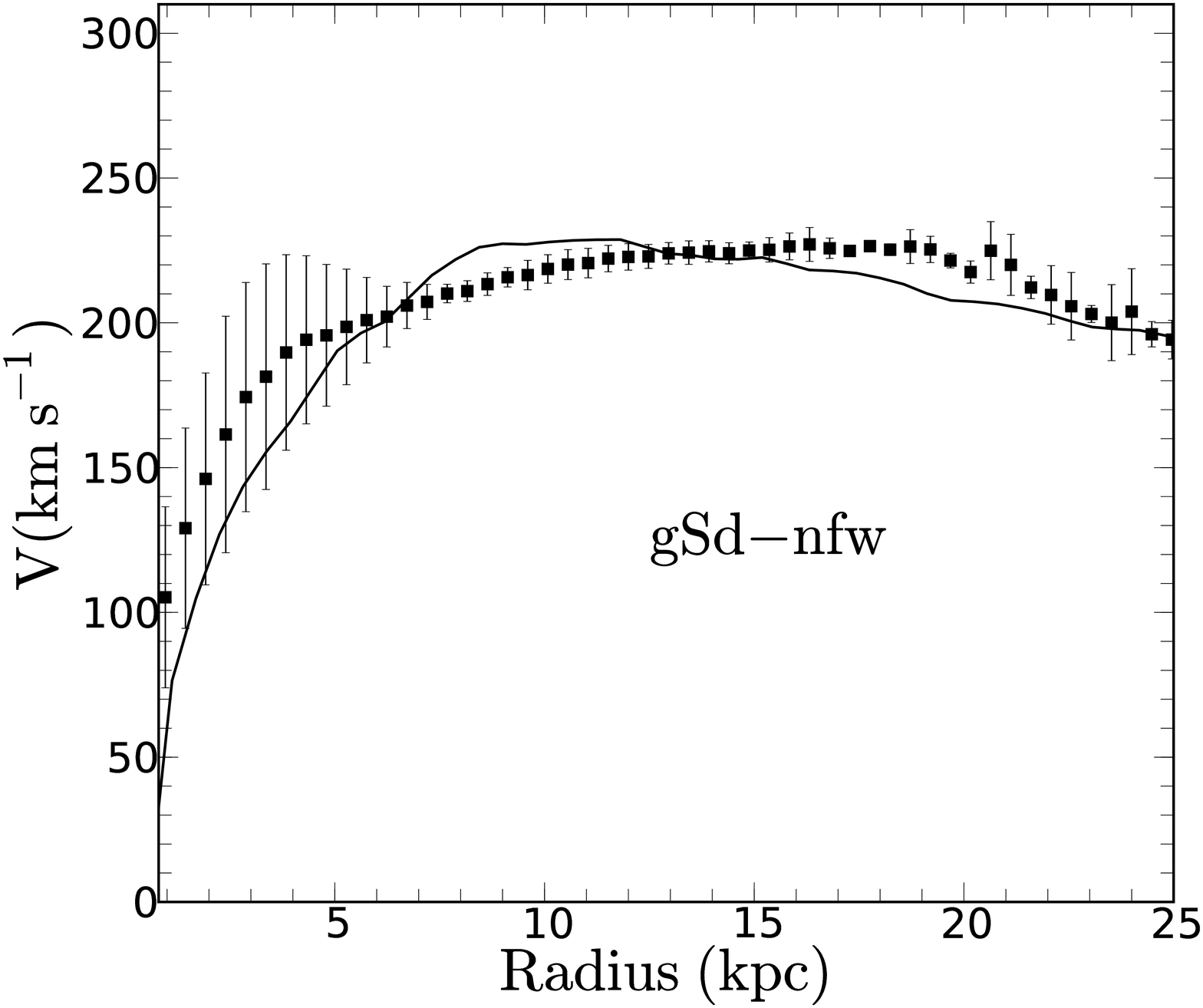}}
                  \end{tabular}
     \caption{This figure shows the comparison between measured and calculated RCs for all the models listed 
in table \ref{tab:simul} for the snapshots with the strongest bar (the epochs are given in table \ref{tab:disp}). Lines and symbols are the same as in figure \ref{fig:vrotstr}.}
     \label{d1}
\end{minipage}
\end{figure*}

The results for the FFT-sticky particles code for three different bar orientations  are shown in figures \ref{stck1}, \ref{stck2} and \ref{stck3} for the gSa, gSb and gSd models respectively. The snapshots were taken at the following epoch T=700 Myrs for gSa, T = 700 Myrs for gSb and T = 900 Myrs for gSd. The intensity maps of the particles, which show the orientations of the bar (MOMENT0 maps) are shown in the left panels and the velocity fields with iso-velocity contours ($\Delta$V = 50 km s$^{-1}$) (MOMENT1 maps) are shown in the middle. The axes of the moment maps are in arcsec. A distance of 16.5  Mpc is adopted to convert from arcsec to parsec scale for all the models. 
Comparison between the expected RCs and the measured  RCs using {\sc rotcur} are shown in the right panels. The expected RCs are presented as continuous lines and the  RCs derived using { \sc rotcur} as black circles. 
     
  {\sc rotcur} over-estimates the rotation when the bar is perpendicular to the major axis except for the gSa model for which most of the mass comes from the bulge component. The top panels of figure \ref{stck1} shows that the measured rotation velocities using { \sc rotcur} and the expected rotation velocities are in good agreement even if the bar is perpendicular with the major axis. This is probably because the higher velocity due to the gas streaming motion along the bar is compensated by the contribution from the bulge. The differences between the measured rotation velocities and those expected from the gravitational potential are the largest when the bar is aligned with the major axis.    
  Figure \ref{fig:vrotstr} shows a comparison between the expected rotation velocities from the total mass of the galaxy and the average rotation velocities (obtained using {\sc rotcur}) for four snapshots. The rotation velocities of 12 simulated galaxies with different bar orientations were extracted and averaged for each snapshot (P.A. varies between 0 and 180 in steps of 15 degrees). The errorbars are the 1-$\sigma$ standard deviations. 
In the top-left panel, the bar starts to form but its effect on the derived rotation velocities is small. The two bottom panels present larger errorbars, with a maximum of 60 km s$^{-1}$ in the inner regions. 
 
The average RCs derived from the snapshots with the strongest bars (see Table 2 for the epochs) for all the models are shown in figure \ref{d1}.  The top panel is the gSa model with an isothermal dark matter halo. The average of the measured RCs seems to be lower compared to the expected rotation velocities calculated from the gravitational potential. The errorbars are larger in the inner parts of the modeled galaxies which implies that the differences between the rotation velocities are only due to the bar orientations. These errorbars represent the variation of the derived RCs compared to the mean RCs. The gSb model (FFT-sticky) is shown in the middle panel. On the left is the model with the isothermal dark matter halo, in the middle the maximum disk model and on the right the NFW model. The gSd model is presented in the bottom panel; the standard model on the left, the maximum disk model in the middle and the NFW model on the right. 

The maximum departure from the average rotation velocities (maximum standard deviation) for each model, are given in table \ref{tab:disp}. The first column is the name of the model, the second column the epoch corresponding to the strongest bar and the last column the maximum deviation from the mean velocity.

%
%
%
\begin{table}
\footnotesize
\caption{Maximum departure from the average velocity (i.e maximum deviation) for the snapshots with the strongest bars.}
\label{tab:disp}
\begin{center}
\begin{tabular}{lccc}
\hline

Model & Epochs & maximum deviation \\
&Myr& km s$^{-1}$ \\
\hline
\\
gSa-st & 800  & 60.72  \\
gSb-st & 1000 & 51.45  \\
gSb-dmx & 600  & 53.45 \\
gSb-nfw & 900  & 52.09 \\
gSd-st & 1000  & 57.27  \\
gSd-dmx & 700  & 68.48 \\
gSd-nfw & 900  &  25.80 \\
\hline
\end{tabular}
\end{center}
\end{table}
\subsection{Bisymmetric model results using DiskFit}

\begin{figure*}
\centerline{
\includegraphics[angle=0,width=19cm]{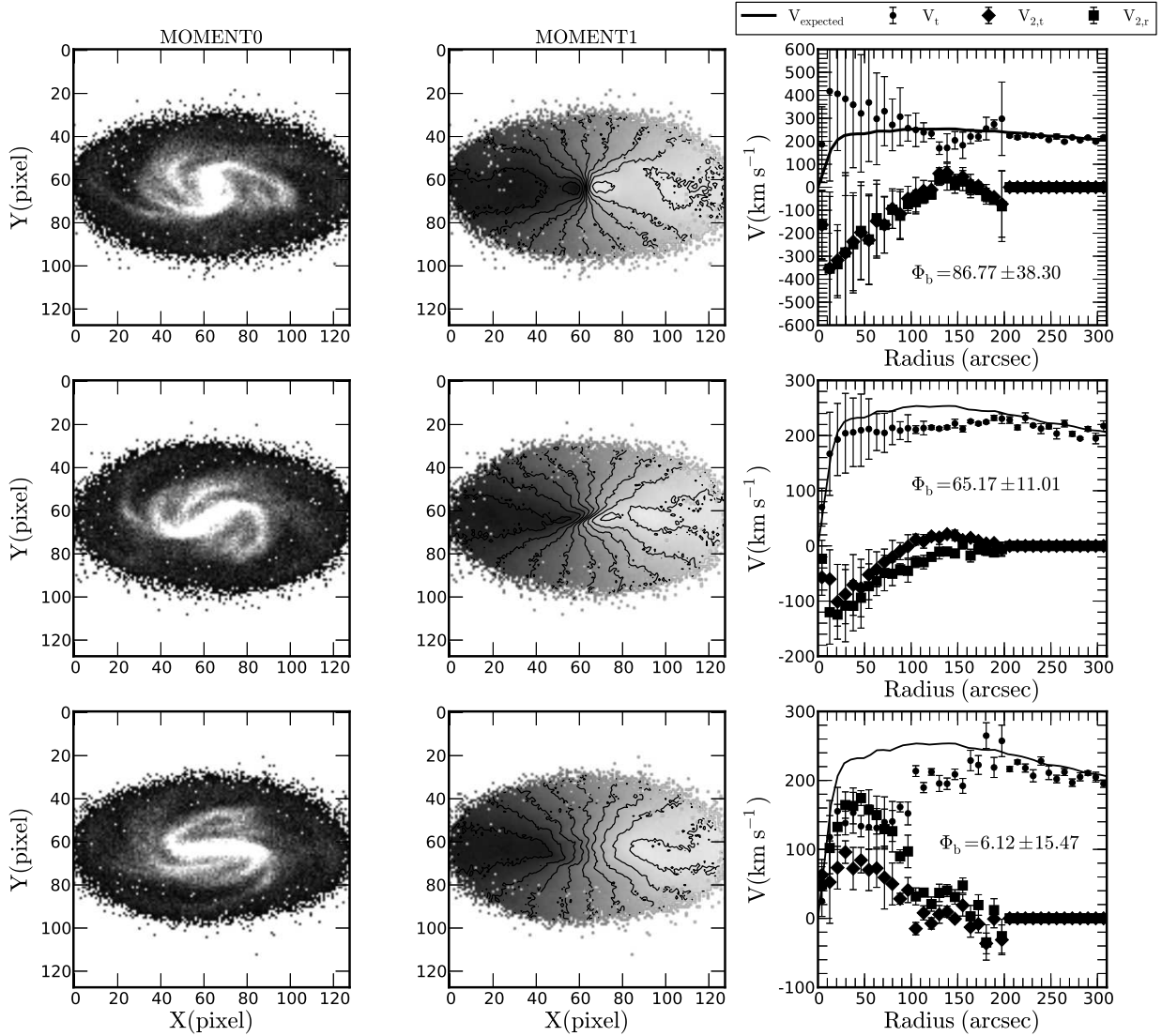}}
\caption{DiskFit results using the bisymmetric model (m=2) for the gSb model (T = 700 Myrs). Top panel: the bar is perpendicular to the major axis, middle panel: intermediate position and bottom panel: parallel to the major axis. 
The first column is the moment0 map, second column the moment 1 map superposed with the iso-velocity contours and the third 
column the comparison between the expected velocities shown as a continuous line and the and the DiskFit results. The mean circular velocity V$_{t}$ is shown as black circles, the tangential component of non-circular motion  V$_{2,t}$ as diamonds, radial components of the non-circular motion as squares and the expected circular velocity V$_{expected}$ as a continuous lines. $\rm \Phi_{b}$ is the bar position angle  in degrees obtained from DiskFit. The axes of the moment maps are in pixels  (1 pixel = 5 arcsec).}
\label{diskor}
\end{figure*}


\begin{figure}
\centerline{
\includegraphics[angle=0,width=9cm]{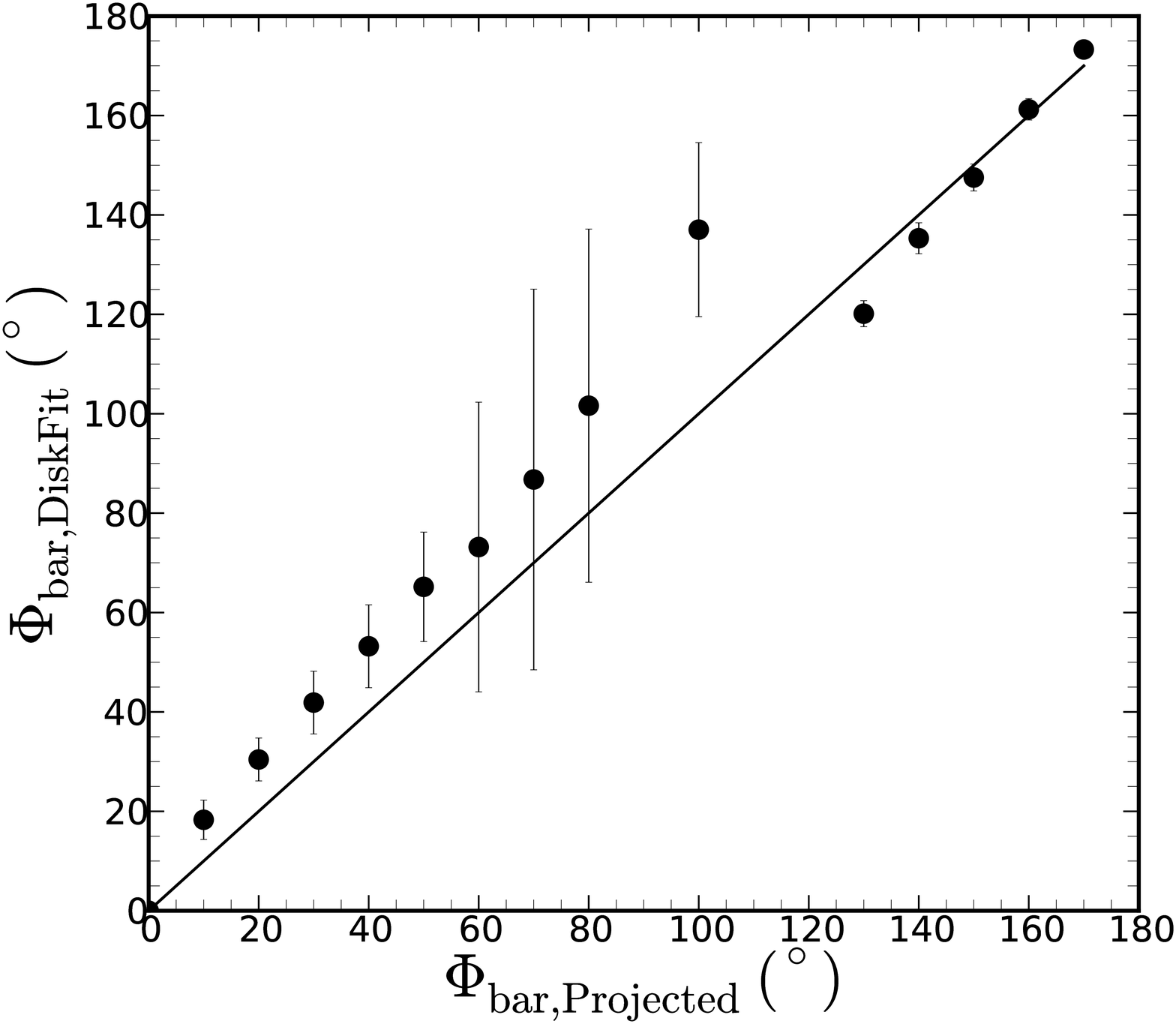}}
\caption{Comparison between the projected bar position angles and those obtained from DiskFit.}
\label{barc}
\end{figure}

  An input parameter file is used to run DiskFit. This file includes the following: a fits file that contains the velocities which are used to derive the kinematics; the initial values for the position angle of the disk and bar, the systemic velocity, the axis ratio and the kinematic center. These parameters can be fixed or let free to vary. For a complete list of parameters the reader is advised to look at the original paper of \cite{2007ApJ...664..204S} and the DiskFit manual. 
    
 The results for the gSb model (T= 700 Myrs) using the bisymmetric model is shown in figure \ref{diskor} for three different bar orientations. Top panel shows the bar perpendicular to the major axis, middle panel shows the bar in an intermediate position and on the bottom panel the bar is parallel to the major axis. The first column is the moment0 map, second column the moment 1 map superposed with the iso-velocity contours and the third column the comparison between the expected velocities shown as a continuous lines and the DiskFit results. The mean circular velocity V$_{t}$ is shown as black circles, the tangential component of non-circular motion  V$_{2,t}$ as blue diamonds, radial components of the non-circular motion V$_{2,r}$ as red squares and the expected circular velocity V$_{expected}$ as a continuous magenta lines. $\rm \Phi_{b}$ is the bar position angle  in degrees obtained from DiskFit. DiskFit is able to reproduce the expected velocities calculated from the gravitational potential and its derivatives when the bar is in intermediate positions but runs into difficulties when the position angle of the bar $\Phi_{b}$ is parallel or perpendicular to the major axis.
This shows that DiskFit produces very large values for the rotation velocities and  the errorbars when the bar is perpendicular with the major axis because of the degeneracy in the model parameters \citep{2010MNRAS.404.1733S}. 

 Figure \ref{barc} compares the projected position angle of the bar $\Phi_{b, projected}$ and those obtained using DiskFit. DiskFit gives higher values as compared to the expected position angle of the bar $\Phi_{b}$ and run into difficulty when the bar is close to 90 degrees.

This is the reason why a different approach is needed when the bar is aligned with the major or minor axis. Here we propose to use N-body simulations of barred galaxies to estimate the non-circular motions.

\section{Test case : NGC 3319}
\label{case}
DiskFit and {\sc rotcur} are both very successful methods for deriving RC
s from velocity maps.  However, Section 3 uses simulations to 
show that 
both methods fail at recovering the true RC
 when a bar is aligned with the major or minor axis of an image.  An alternate approach may be to use simulations to correct the RCs for non-circular motions.  This will avoid many of the issues with either {\sc rotcur} or DiskFit.  We illustrate this method using NGC 3319, which has its bar aligned with the major axis, as 
a test case. NGC 3319 is a gas-rich barred spiral galaxy classified as SBcd(rs) by \cite{1991rc3..book.....D} at a distance of 14.3 Mpc \citep{1999ApJ...523..540S}.   We compare the inferred mass profile for the uncorrected and corrected RCs and show that the conclusions about the dark halo structure vary greatly.  While our method still needs refinement, this result clearly demonstrates the importance 
of taking into account non-circular motions.

\subsection {Observations and data reduction}
\begin{figure*}
\centering
\begin{minipage}{140mm}
   \begin{center}
  \begin{tabular}[t]{c}

         \hspace*{.0em} \vspace{.0em}
           \subfigure{\includegraphics[width=0.54\linewidth]{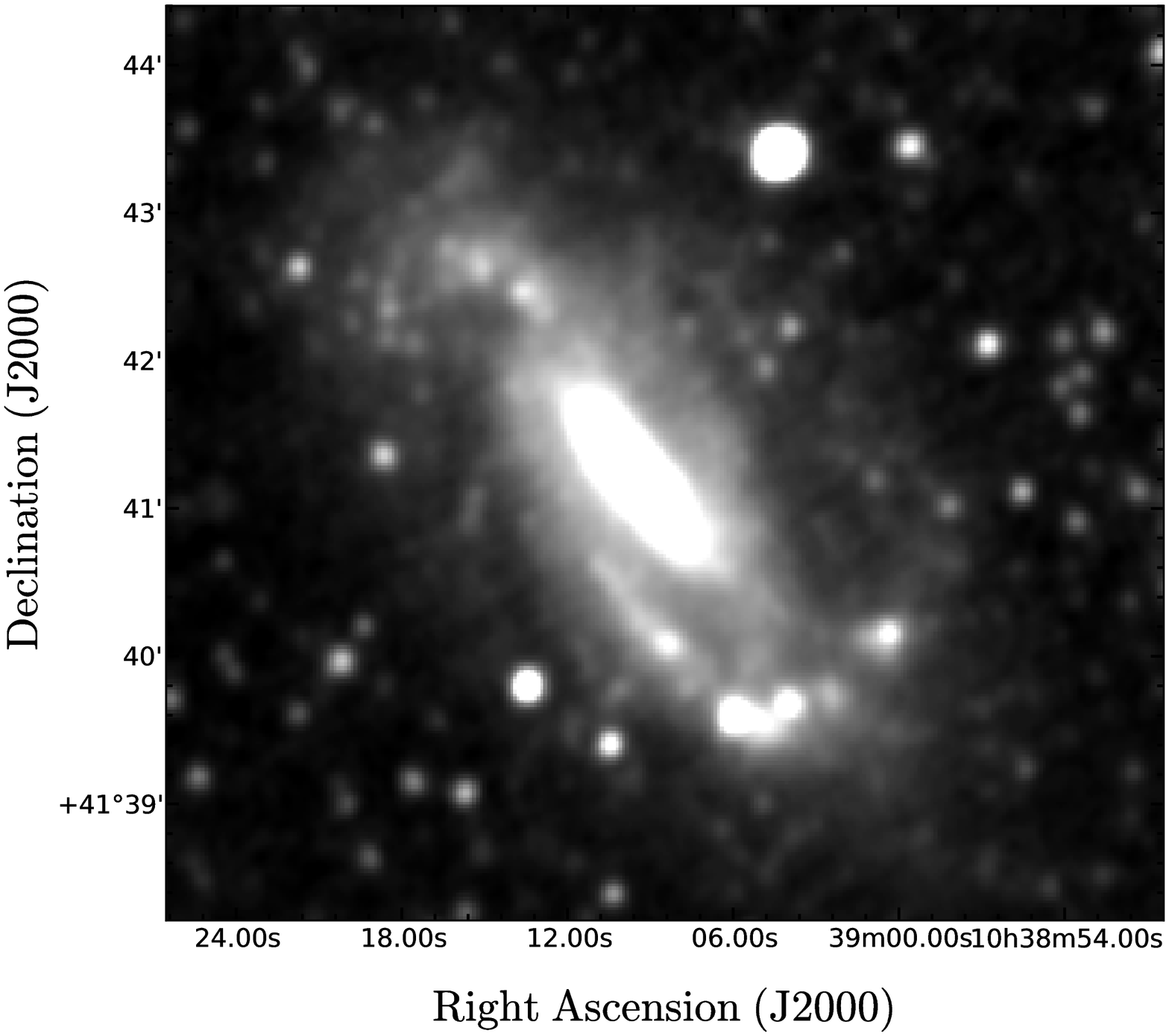}}
	   	\hspace*{1.50em} \vspace{.0em}
        \subfigure{\includegraphics[width=0.54\linewidth]{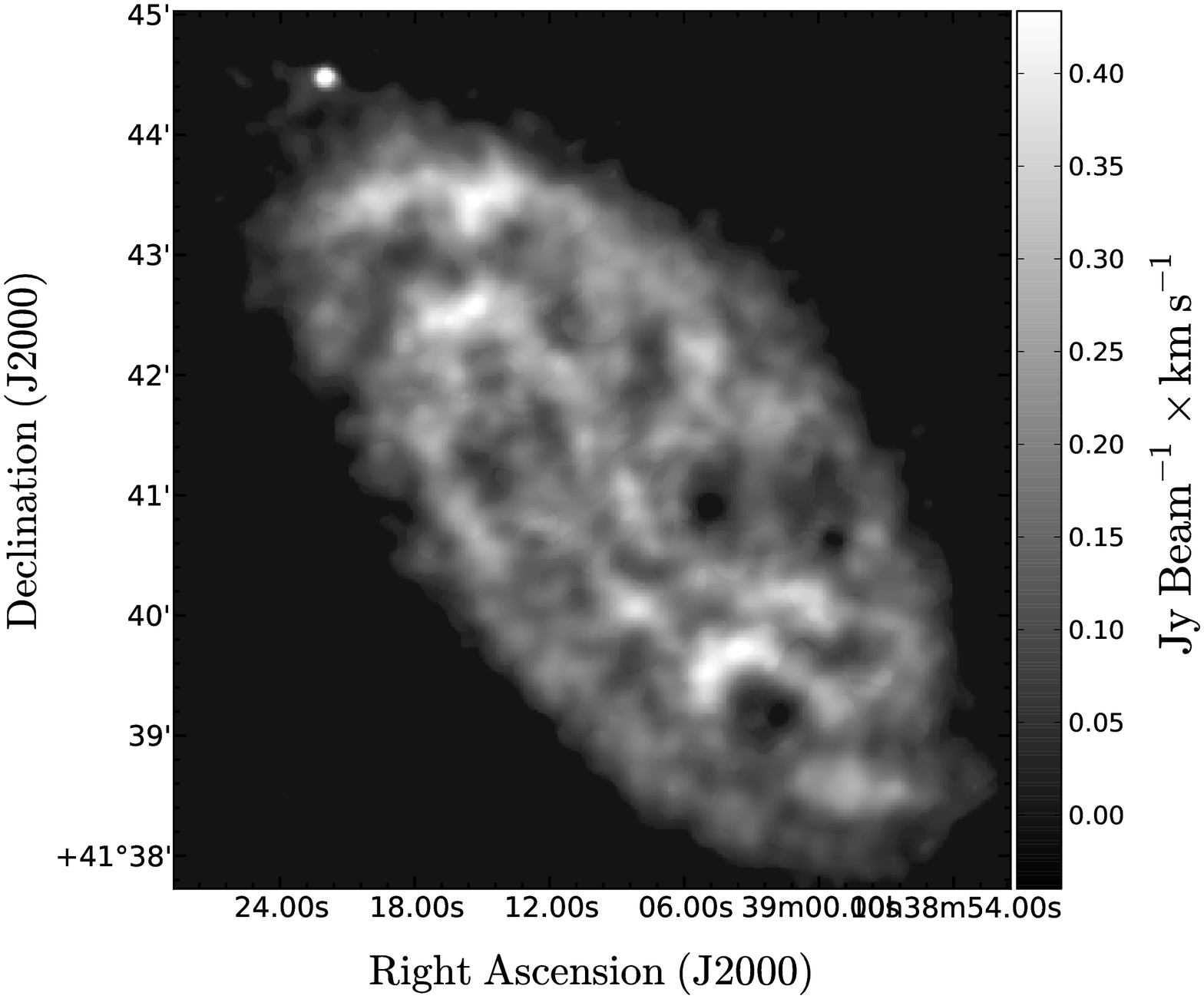}} \\
         \hspace*{.0em}
        \subfigure{\includegraphics[width=0.55\linewidth]{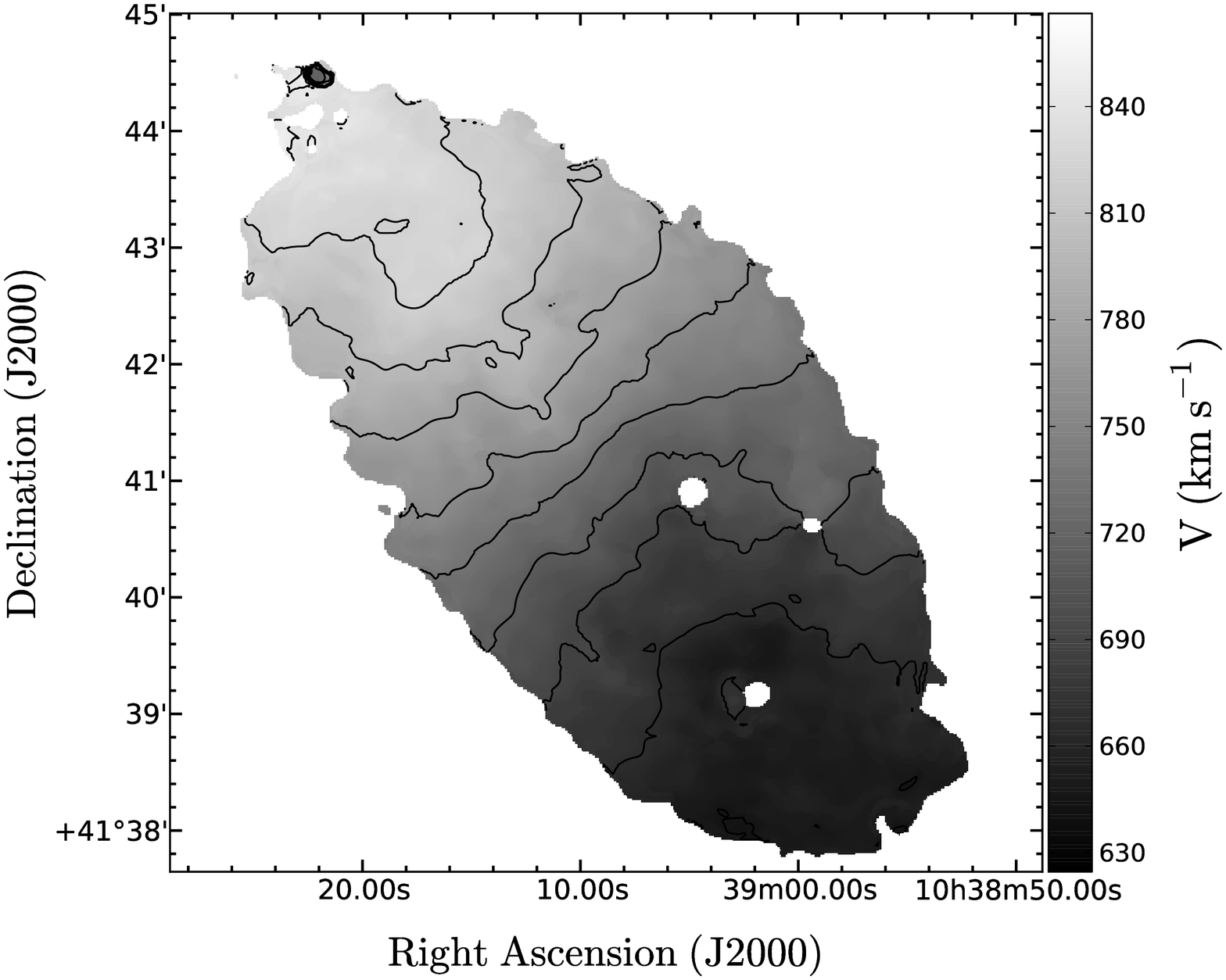}} 
        \hspace*{.0em} \vspace{.0em}
          \subfigure{\includegraphics[width=0.55\linewidth]{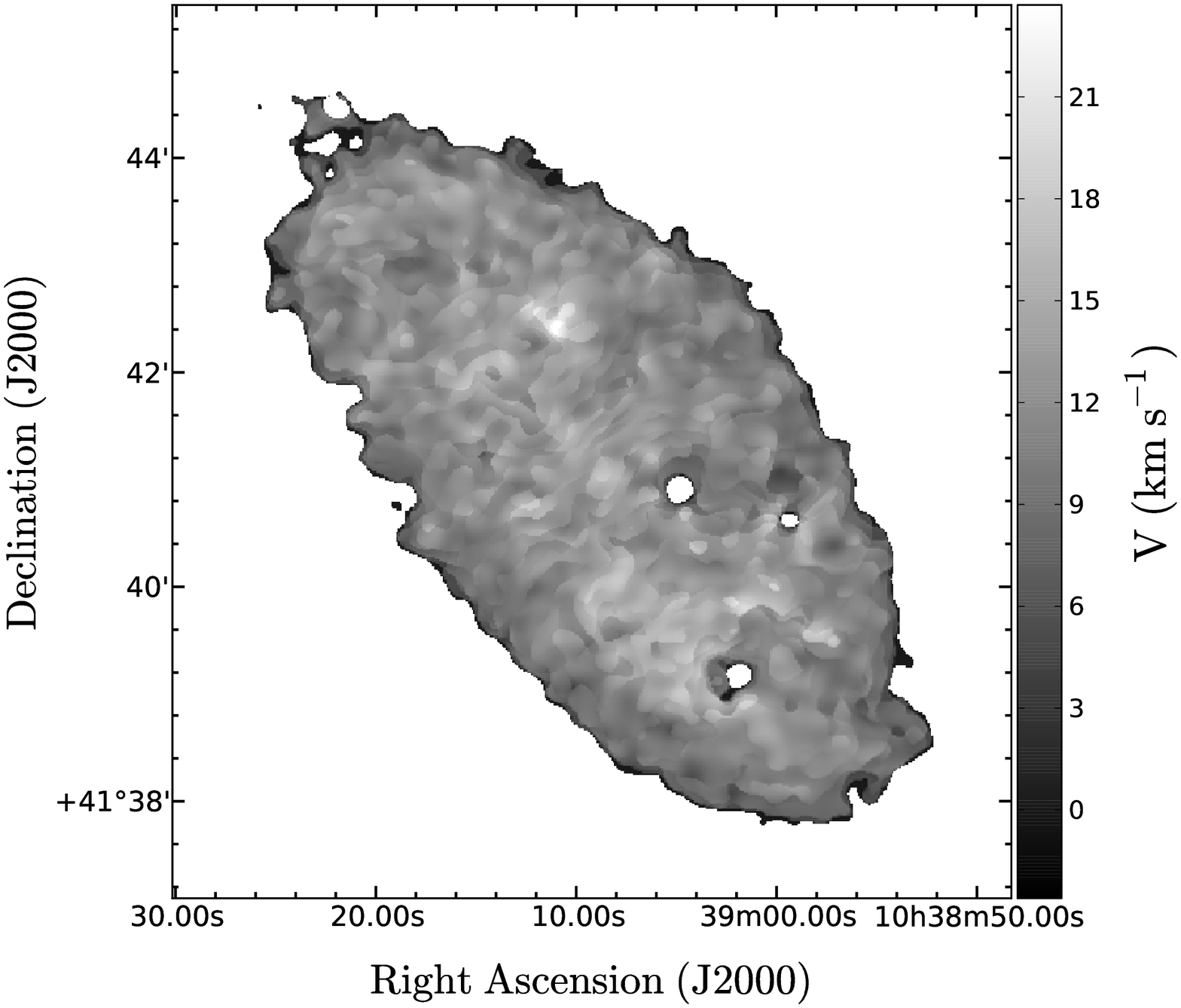}}\\
	   	\hspace*{.0em} \vspace{0.0em}
                       \end{tabular}
     \caption{This figure shows several maps of NGC 3319. The top left panel shows the WISE W1 band (3.4 microns) image, where the bar and spiral arms are clearly seen; the moment0 map (H{\sc i} distribution) is shown in the top right panel, the moment1 map (H{\sc i}  velocity field) 
in the bottom left panel and the moment2 map (H{\sc i} velocity dispersion) in the bottom right corner. }
     \label{maps}
  \end{center}
\end{minipage}
\end{figure*}

The H{\sc i} observations were obtained using the C configuration of the Very Large Array (VLA) for a total observing time of 5.75h on-source. Details of the observations are presented in \cite{1998MNRAS.294..353M}. The raw data were retrieved from the archive, edited and calibrated using the standard NRAO Astronomical Image Processing System (AIPS) packages. Data cubes were then obtained and cleaned with the AIPS task {\sc imagr} using robust0 weighting, which optimizes sensitivity and spatial resolution.
1031+561 is the phase and amplitude calibrator and 3C286 the flux calibrator. The C array data provide the proper spatial resolution for mass model analysis. The data cubes have 64 channels with 10.33 km s$^{-1}$ velocity resolution and 16 arcsecs angular resolution, which corresponds to a linear physical resolution of $\sim$1 kpc at the adopted distance of 14.3 Mpc. The moment maps were extracted from the data cubes using the AIPS task {\sc momnt}. Figure \ref{maps} shows four different maps for NGC 3319. The WISE W1 image is shown at the top left. This band is essentially free from dust extinction and gives the best representation of the stellar content of the galaxy. The moment 0 map (H{\sc i} gas distribution) is at the top right, the moment1 map (velocity field)  at the bottom left  and the moment2 map (velocity dispersion) in the bottom right panel.

\begin{figure}
\centerline{
\includegraphics[angle=0,width=10cm]{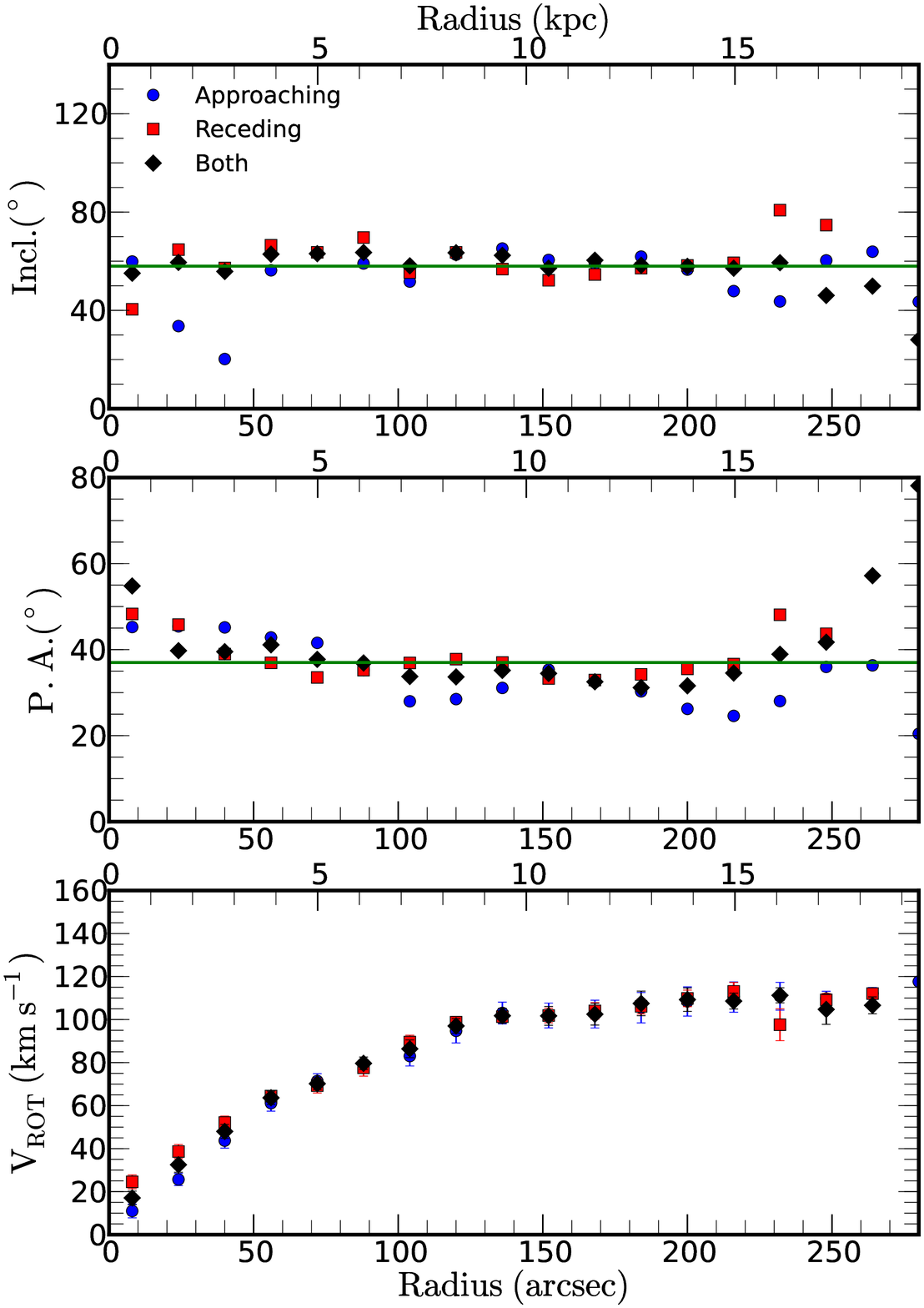}}
\caption{Tilted ring model results obtained with the GIPSY task {\sc rotcur}. The variation of the inclination as a function of radius is shown in the top panel and the position angle (PA) variation in the middle panel. The RC presented in the bottom panel was obtained with all the kinematical parameters, including PA and the inclination, kept fixed to their average values.
 The results from the receding side are presented as red squares, those from the approaching side as blue circles and from both sides as black diamonds. The lines correspond to PA= 37$^{\rm o}$ and i = 58$^{\rm o}$.}
\label{tilt}
\end{figure}

The tilted ring method, implemented in the GIPSY task {\sc rotcur}, is used to derive the RC
. The moment1 velocity map was used as input for {\sc rotcur}. 
The tilted ring analysis was done in three steps:
\begin{itemize}
\item First, the position angle (P.A.) and inclination angle ($i$) were fixed using the values of 37$^{\rm o}$ and 58$^{\rm o}$ found by \cite{1998MNRAS.294..353M} in order to estimate the systemic velocity and the kinematical centre. Our measured rotation centre and systemic velocity of 742 $\pm$ 2 km s$^{-1}$  are consistent with \cite{1998MNRAS.294..353M} (see their table 3). 
\item Second, the systemic velocity and the rotation centre were kept fixed using the values obtained in the first step and the position angle and inclination were allowed to vary to investigate their variation as a function of radius.
\item Finally in the final step, the kinematical parameters were all fixed and only the rotation velocities were derived.
\end{itemize}
The results from the tilted ring analysis are presented in figure \ref{tilt}. The variation of the inclination as a function of radius is shown in the top panel, the position angle in the middle panel and the RCs in the bottom panel. The results from the receding half is presented as red squares, those from the approaching half as blues circles and those obtained using both halves as black diamonds.

%
%

\subsection {Model-based correction to the RC}
\label{corr}
As discussed in section \ref{res}, the RC is under-estimated when the bar is aligned with the major axis because of the gas streaming along the bar. 
Therefore, the effect of those non-circular motions need to be estimated, and a correction needs to be applied to the RC before it can be used for the mass models.

 We introduce the following step to correct the RCs for the non-circular motion induced by the bar:
\begin{itemize}
  \item The correction is derived from the differences between the expected RC of the model, which is calculated from the true gravitational potential of the modeled galaxy, and the observed RC measured using the tilted ring method ({\sc rotcur})
  \item The correction is scaled in terms of velocity using the $V_{max}$ ratio, i.e. ($\frac{V_{max, object}}{V_{max, model}}$) and in terms of radius using the ratio between the disk scale lengths, i.e. ($\frac{R_{d, object}}{R_{d, model}}$)  before being used to correct the RC of an actual galaxy.
  \item The modeled galaxy (model) and the actual galaxy (object) should have similar bar properties, the distance also have to be taken into account.
\end{itemize}
The correction is done using the following
\begin{equation}
V_{corrected} = V_{uncorrected} + correction\times\frac{V_{max, object}}{V_{max, model}}
\end{equation}
where $ \rm correction=V_{obs, model} - V_{expected}$, V$_{obs, model}$ is the observed velocities obtain from {\sc rotcur}, V$_{expected}$ the expected circular velocites; V$_{corrected}$ and V$_{uncorrected}$ are the corrected and uncorrected rotation velocities of NGC 3319 respectively.
\\

\begin{figure}
\centerline{
\includegraphics[angle=0,width=9cm]{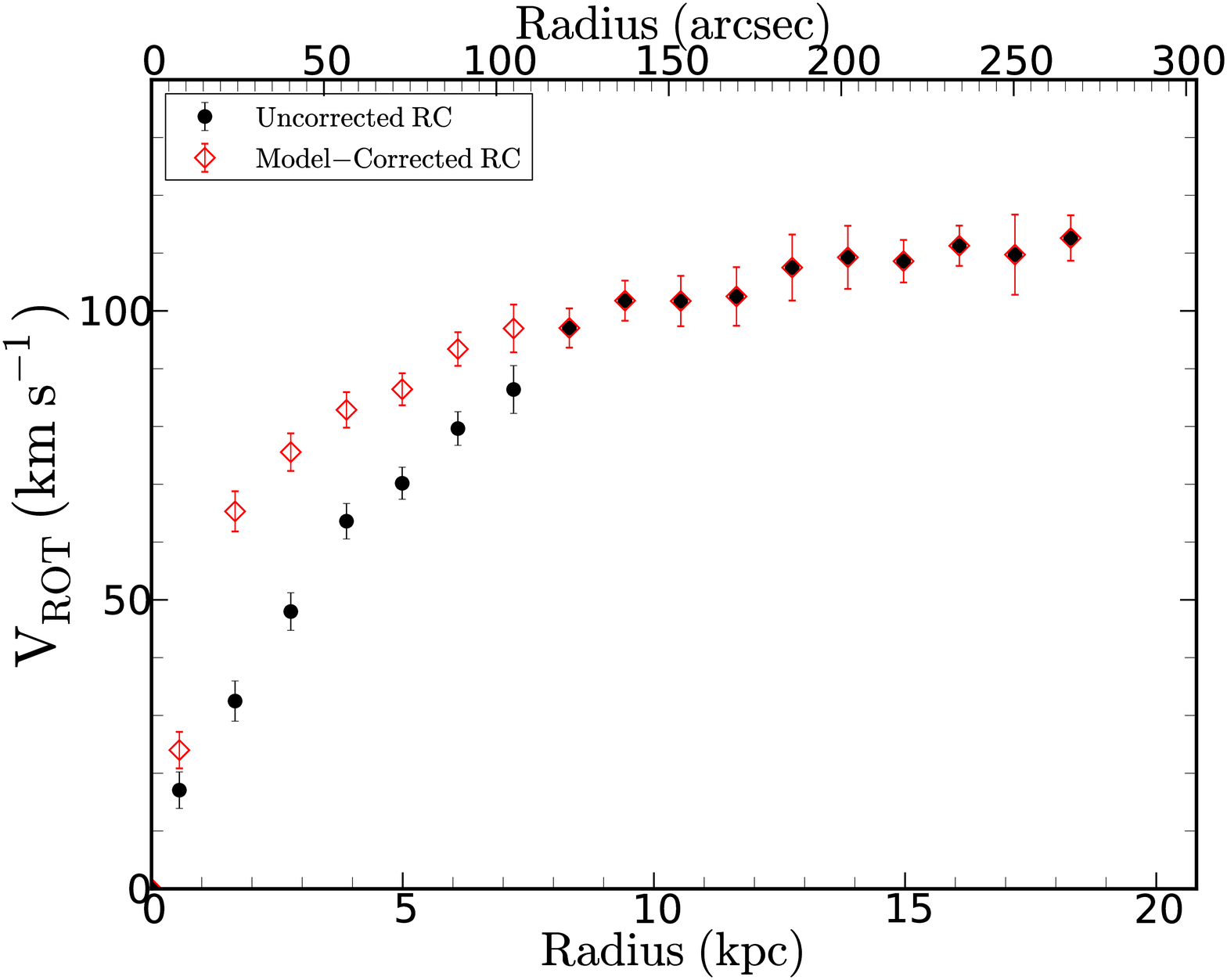}}
\caption{Comparison between the model-corrected and uncorrected RCs.}
\label{rcs}
\end{figure}

 To apply the corrections, the data from the model were normalized by the optical scale length in radius and by the maximum velocity in velocities using the steps described above. The data is then extrapolated to obtain the correction at a given radius. The distance of NGC 3319 is used to convert from angular scale into physical scale.
Comparison between the corrected and uncorrected RC is shown in figure \ref{rcs}.

Despite the fact that NGC 3319 is classified as a SBcd galaxy, the gSb model presented at the bottom panel of figure \ref{stck2} is used to derive the corrections. This is because the gSd model does not have a bulge component (see table \ref{tab:simul}) and that the bulge-to-disk (B/D) ratio of the gSb model obtained from the ratio between the bulge mass and disk mass given in table \ref{tab:simul} M$_{b}$/M$_{d}$ = 0.239 is very similar to the B/D = 0.24 of NGC 3319 estimated from the 3.6 micron surface brightness profile.

In this illustrated test case, we do not pretend to accurately correct for the non-circular motions induced by the bar. In the continuation of this work, we intend to match even more accurately the simulated galaxies to the actual galaxies. However, we think that the gSb simulation used is close enough to NGC 3319 to show clearly how not taking into account the effect of bars can lead to deriving very different parameters for the derived dark matter component. 
We still have to investigate if it is the best way to scale the model to the actual galaxies. However, this is surely close enough to serve the purpose of this study.

\subsection {Mass model}
\subsubsection{Dark Matter Halo component}
The distribution of DM is characterized by a density profile. The observationally motivated pseudo-isothermal halo (ISO) and the cosmologically motivated Navarro, Frenk and White (NFW) \citep{nfw96} profiles are widely used.
The mass model is done by comparing the observed RCs with the quadratic sum of the luminous and dark matter components using a chi-squared minimization technique.  The mass models were carried out using the GIPSY tasks {\sc rotmas} and {\sc rotmod}.

The RC is given by:
\begin{equation}
\rm V_{rot}^{2} = V_{gas}^{2} + V_{*}^{2} + V_{halo}^{2}
\end{equation}
where $V_{gas}$ is the gas contribution, $ V_{*} $ the stars contribution and $V_{halo}$ the contribution of the dark matter component.

\subsubsection*{The pseudo-Isothermal (ISO) Dark Matter (DM) Halo Model}

The pseudo-isothermal (ISO) dark matter halo is a  core-dominated type of halo. The ISO density profile is given by:
\begin{equation}
\rm \rho_{ISO}(R)={\rho_0 \over 1+({R\over R_c})^2}
\label{eq:ro_iso}
\end{equation}
where $\rho_0$ is the central density and $R_c$  the core radius of  the halo. The velocity  contribution of a ISO halo is given by: 
\begin{equation}
\rm V_{ISO}(R)=\sqrt{4\pi~ G~\rho_0~ R_c^2(1-{R\over R_c}~ atan ({R\over R_c}))}
\label{eq:v_iso}
\end{equation}

We can describe the steepness of the inner slope of the mass density profile with a power law $\rho \sim r^\alpha$.
In the case of the ISO halo, where the inner density is an almost constant density core, $\alpha = 0$.

\subsubsection*{The NFW DM Halo Model}

The NFW DM halo profile was derived from cosmological simulations and it is commonly accepted in the $\Lambda$CDM framework. The NFW halo density  profile is described by: 

\begin{equation}
\rm \rho_{NFW}(R)={\rho_i \over {R\over R_S} (1+{R\over R_S})^2}
\label{eq:ro_nfw}
\end{equation}
where $\rho_{i} \approx 3H_{0}^{2}/(8 \pi G)$ is the critical density for closure of
the universe and $R_{S}$ is a scale radius. The  velocity contribution corresponding to this halo  is given by: 
\begin{equation}
\rm V_{NFW}(R)=V_{200}~\sqrt{{ln(1+cx)-cx/(1+cx) \over x(ln(1+c) -c/(1+c))]}}
\label{eq:v_nfw}
\end{equation}
 where $\rm V_{200}$ is  the velocity at the virial radius $\rm R_{200}$,   
  $\rm c = R_{200}/R_S$ gives the concentration parameter of the halo and  x is defined as $\rm R/R_S$. The relation between $\rm V_{200}$ and $\rm R_{200}$ is given by:
  
  \begin{equation}
\rm V_{200}= {R_{200} \times H_0 \over 100}
\label{eq:v200}
\end{equation}
where $\rm H_{0}$ is the Hubble constant taken as  $\rm H_{0}=72.0 \ km \ s^{-1} \  Mpc^{-1}$ \citep{2009ApJS..180..225H}.

\subsubsection{Luminous matter Components}
 
\subsubsection*{Gas Contribution}

The contribution from the gas is estimated from the moment0 maps. The task {\sc ellint} is used to compute the H{\sc i} gas density profile  using the kinematic parameters obtained with {\sc rotcur}. {\sc ellint} divides the moment0 map into concentrated rings and estimates the density for each ring.
The output from {\sc ellint} is then used in  {\sc rotmod} to calculate the gas contribution in the mass model, assuming an infinitely thin disk.
The profile was multiplied by 4/3 to take into account the primordial Helium contribution.

\subsubsection*{Stellar Contribution}
\label{sec:stellar}

The SPITZER 3.6 microns surface brightness profile is used to calculate the stellar contribution. Near-Infrared probes most of the emission from the old stellar disk population. It is also less affected by dust and therefore represents the bulk of the stellar mass.  The data were retrieved from the SPITZER archive. After  the foreground stars were removed, the images were fitted with concentric ellipses using the {\sc ellint} task in GIPSY. The results were then converted into a surface brightness profile in mag/arcsec$^{2}$ using the method in \cite{2008AJ....136.2761O}.

The surface brightness profile is corrected for inclination and galactic extinction before being converted into mass density.
 The method used by \cite{2008AJ....136.2761O} was also adopted to convert the luminosity profile into a mass density profile.
 The density profile is given by:
\begin{equation}
\rm \Sigma[M_{\odot} pc^{-2}] = \Upsilon^{3.6}_{*} \times 10^{-0.4 \times (\mu_{3.6} - C^{3.6})}
\end{equation}
where  $\Upsilon^{3.6}_{*}$ is the stellar mass-to-light ratio in the 3.6 microns band, $\mu_{3.6}$ the surface brightness profile and C$^{3.6}$ is a constant used for the conversion from mag/arcsec$^{2}$ to L$_{\odot}$/pc$^{2}$.
C$^{3.6}$ is given by:

\begin{equation}
\rm C^{3.6} = M_{\odot}^{3.6} + 21.56
\end{equation}
where M$_{\odot}^{3.6}$ is the absolute magnitude of the Sun in the 3.6 microns band.
Using the distance modulus formula and the distance to the Sun, \cite{2008AJ....136.2761O} found:
\begin{equation}
\rm M_{\odot}^{3.6} = m_{\odot}^{3.6} + 31.53 = 3.24
\end{equation}
 
 \subsubsection*{The bulge-disk decomposition}
 
\begin{figure}
\centerline{
\includegraphics[angle=0,width=9cm]{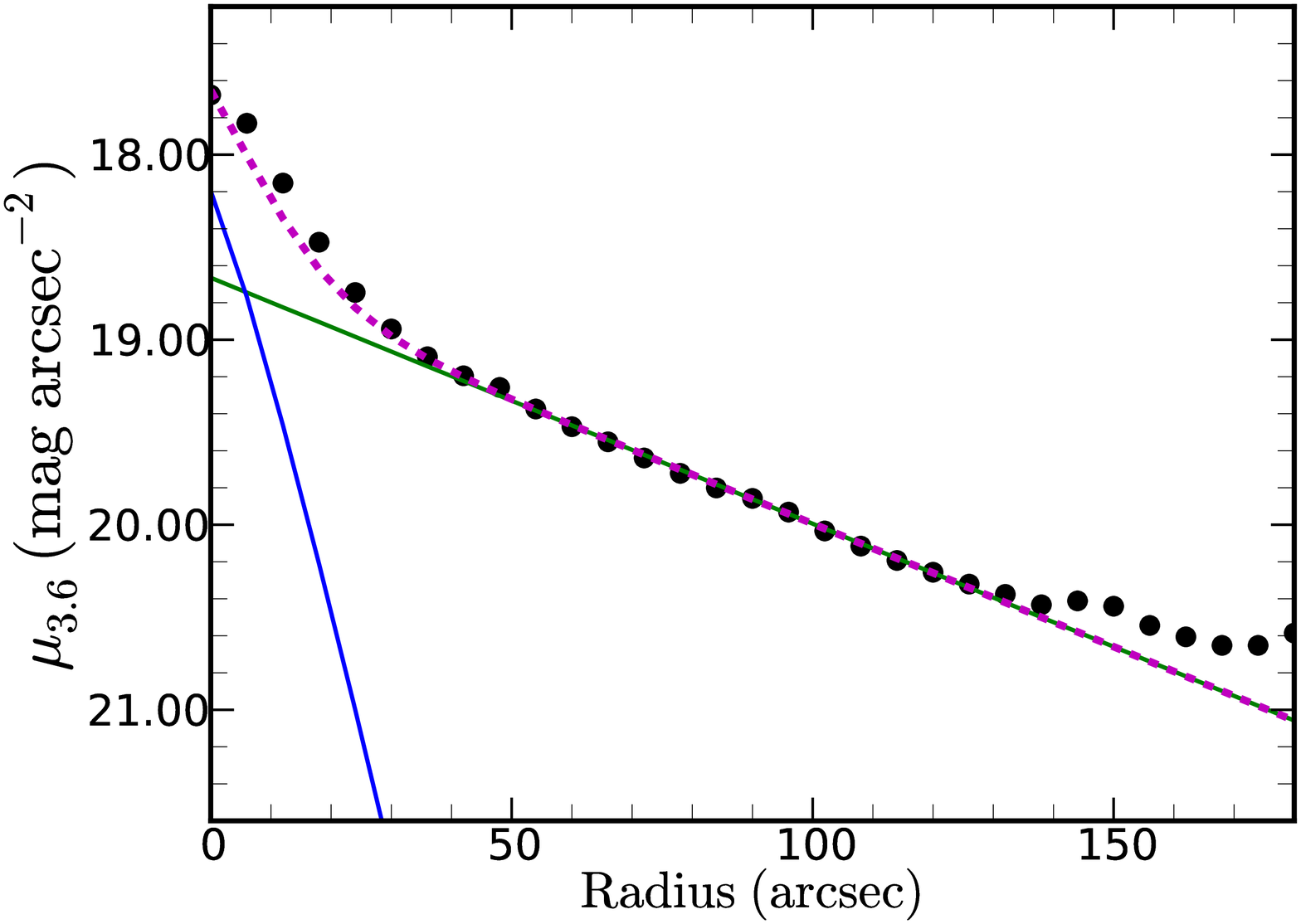}}
\caption{Bulge-disk decomposition, the green line is the best fit for the disk, the blue for the bulge and the dashed magenta line the sum of the contribution from both components.}
\label{bulge}
\end{figure}

 The surface brightness profile in figure \ref{bulge} clearly shows that this galaxy has a bulge in the center, thus a bulge-disk decomposition is needed to better constraint the contribution from the stellar component.
 
The disk component was fitted with an exponential disk using the following equation

\begin{equation}
\mu (r) = \mu_{0} + 1.10857\frac{R}{R_{d}},
\end{equation}

where R$_{d}$ is the disk scale length and $\mu_{0}$ the central surface brightness. The remaining profile is fitted with a Sersic profile by letting the Sersic index n, the effective radius R$_{e}$ and the effective surface brightness $\mu_{e}$ as free parameters. 
The bulge is therefore described by

\begin{equation}
\mu (r) = \mu_{e} + 2.5b_{n}[(\frac{R}{R_{e}})^{1/n} + 1],
\end{equation}
where b$_{n}$ = 1.9992n - 0.3271 for 0.5 < n < 10 \citep{1989woga.conf..208C}
The best fit result shown in figure \ref{bulge} gives R$_{d}$ = 5.77 $\pm$ 0.06 kpc, $\mu_{0}$ = 18.66 $\pm$ 0.02 mag for the disk and n = 0.87 $\pm$ 0.02, R$_{e}$ = 2.04 $\pm$ 0.05 kpc and $\mu_{e}$ = 21.74 $\pm$ 0.31 mag for the bulge. 

 \subsubsection*{Mass-to-light ratio}
 
 The mass-to-light ratio ($\Upsilon_{*}$) for the IRAC1 band is normally derived using the J-K color (see \citealt{2008AJ....136.2761O}, \citealt{de-Blok:2008oq}). However, this method gives a higher  mass-to-light ratio for the disk component than the bulge component which is unphysical. \cite{2013ApJ...773..173D} uses chemo-spectrophotometric galactic evolution (CSPE) models to constrain the mass-to-light ratio of nearby galaxies as function of radius. We use their results to estimate the mass-to-light of NGC 3319.  NGC 3319 and NGC 2403 have the same morphological type (SBcd) and similar maximum velocity but NGC 3319 has a larger bulge compared to NGC 2403. The values used for the mass model are given in table \ref{mas}.
 
%
%

\subsection{Results}
\begin{figure*}
\centerline{
\includegraphics[angle=0,width=18cm]{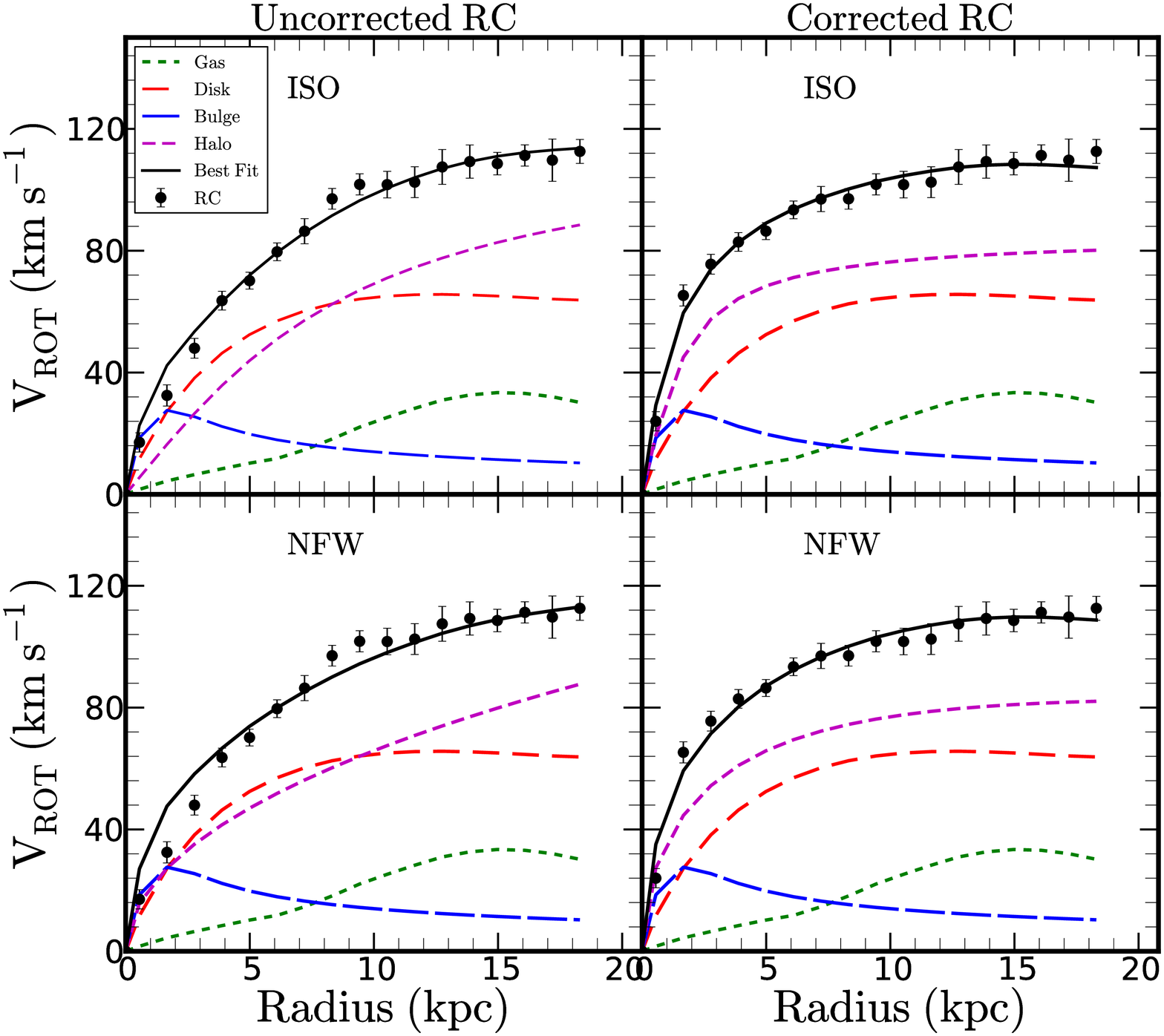}}
\caption{Mass models using ISO DM halo model (top) and NFW DM halo model (bottom) for the uncorrected (left) and model-corrected (right) RCs. Lines and symbols are shown on the top left corner of the first panel.}
\label{nfw}
\end{figure*}

The mass model results for the uncorrected and corrected RC
s are presented in figure  \ref{nfw} for the ISO and NFW models. The observed RC is shown as filled black circles with error bars, the gas contribution as dashed green lines, the stellar disk contribution as dashed red lines, the stellar bulge contribution as long-dashed blue lines and the contribution from the dark matter halos as dashed magenta lines. The continuous lines are the best-fit models to the RC. Table \ref{mas} summarizes the results for the ISO and NFW models. The columns are described as follows, the first column is the type of DM halo used, the second column gives the parameters of the fits, the third and fourth columns are the results.
For the ISO models, the dark matter halo is much more centrally concentrated when the corrected RC
 is used with a R$_{C}$ $\sim 1$ kpc compared to $\sim 5$ kpc for  the uncorrected RC
 and  a much larger $\rho_{0}$ (see table \ref{mas} for details). It is clear that the corrected and uncorrected RC give very different results for the dark matter parameters which shows the importance of those corrections for non-circular motions. The same is true for
the NFW models where the concentration index c is five times larger when the corrected RC is used and R$_{200}$ much smaller. Both the ISO and NFW models gives very good fits to the corrected RC with smaller $\chi^{2}_{r}$. This confirms that the parameters of the dark matter halo are strongly affected by the non-circular motions induced by the bar, especially when the bar is aligned with the major axis. A large portion of the mass will be missed if the RC
 is not corrected and different conclusions could be drawn from the dark matter parameters. 
The difference between the uncorrected and corrected RC is still significant even after the bulge component has been taken into account. This is a clear demonstration that corrections must be applied to the observed RC so that it really represents the dynamical mass of barred systems. 


\begin{table}
\caption{Results for the ISO and NFW dark matter halo models using the uncorrected and corrected RCs.}
\begin{center}
\begin{tabular}{c c  c  c }
\hline
\\
Halo Model&Params &Uncorrected RC&Corrected RC\\
\hline 
ISO& $\rho_{0}$& 5.55$\pm$0.99 & 74.08$\pm$15.15\\
 & R$_{c}$& 6.91$\pm$1.11&1.34$\pm$0.16\\
 &$\Upsilon_{d}$& 0.25&0.25 \\
&$\Upsilon_{b}$& 0.40&0.40 \\
&$\chi^{2}_{r}$&1.35&0.75\\\\
NFW& c& 0.43$\pm$0.01&5.18$\pm$0.73\\
 & R$_{200}$&  147.25$\pm$11.00&56.46$\pm$3.44\\
 &$\Upsilon_{d}$&0.25 &0.25 \\
&$\Upsilon_{b}$&0.40&0.40 \\
&$\chi^{2}_{r}$&3.24&1.44\\

\hline

\end{tabular}

\addtocounter{footnote}{-2}
\end{center}
\footnotesize{{\it Notes}:  $\Upsilon_{d}$ is the mass-to-light ratio for the stellar disk and $\Upsilon_{b}$ for the bulge. The central DM density $\rho_{0}$ is given in units of 10$^{-3}$ M$_{\odot}$ pc$^{-3}$; R$_{c}$ and R$_{200}$ are in kpc.}\\

\label{mas}
\end{table}

%

\section{Summary}
\label{disc}

We have presented the analysis of snapshot observations of N-body/hydrodynamic simulations of barred spiral galaxies with different bar strengths and orientations to quantify the non-circular motions induced by a bar. .
In the simulations, the bar forms naturally due to gravitational instabilities. Therefore, snapshots of simulated galaxies spanning a wide range of bar strengths and bar orientations can be extracted and analyzed. 
Several models with different morphological types and dark matter halo distributions are analysed. 

The results from the tilted ring method using {\sc rotcur} show that the rotational velocities are largely under-estimated if the bar is parallel to the major axis and over-estimated if the bar is perpendicular except for an Sa galaxy which has an important central bulge for both simulations. This is consistent with \cite{2008MNRAS.385..553D}. This corresponds to the elongated orbits of the gas along the bar, which enhance the velocities at pericenter and reduce them at apocenter. Part of the gas and dust are moving alongside the bar which increases the measured circular velocity when the bar is parallel to the major axis and decreases it if the bar is aligned with the minor axis. The effect is smaller when the bar is in an intermediate position.

We also use DiskFit for our analysis. DiskFit is specifically designed to fit non-axisymmetric distortions from 2D velocity maps \citep{2007ApJ...664..204S}. 
However, DiskFit works only when the bar is not aligned with the symmetry axes of the projected galaxy, and strong distortions are seen in the velocity field.

\cite{2010MNRAS.404.1733S} noted that the circular velocity is biased low when the bar is parallel to the major axis (apocentre) since the gas follows an elliptical pattern and that the circular speed should be higher when the bar is aligned with the minor axis. 
%

The test case in section \ref{case} shows that the distribution of mass in NGC 3319 is very affected by the non-circular motions induced by the bar. For example, the central dark matter core is much more compact when the corrected RC is used. The ISO and NFW results are very different in all cases for the corrected and uncorrected RCs. 

For future work, snapshot observations with different bar strengths and orientations will be studied to estimate the corrections that need to be applied to  the observed rotation velocities of a sample of barred galaxies with a variety of bar strength and bar orientation. A sample of barred galaxies with different bar sizes and orientations will be selected and modeled. This will help us to compare the results from the observations and the simulations for the individual galaxies.The main aim of this future work is to try to match as closely as possible the parameters of the model and the parameters of the galaxies, especially those of the bar. In particular, we will try to closely match the DM halo-to-luminous disk ratios.

%
%
%

\section*{Acknowledgments}
We thank the anonymous referees for the comments and suggestions. CC's work is based upon research supported by the South African Research Chairs Initiative (SARChI) of the Department of Science and Technology (DST), the SKA SA and the National Research Foundation (NRF); ND and TR's work are supported by a SARChI's South African SKA Fellowship. We thank T. Jarrett for the Wise image.

\appendix
\section*{APPENDIX}
\section{Numerical details about the simulations}
\label{apa}
The first simulation was done 
with a 3D Particle-Mesh code, based on FFT, with a useful grid of
128$^3$. The algorithm of \cite{james1977} is used to suppress
the Fourier images, and gain a factor 8 in memory.
A total number of particles of 240 000 is distributed
on the grid, and the density is computed with the cloud
in cell algorithm (CIC).
The gas is represented by sticky particles,
colliding with inelastic collisions. At each encounter, the relative
velocity of the clouds is multiplied by 0.65 in the radial direction, and 1 in the 
tangential direction, which keeps the angular momentum conserved.

The star formation is assumed to follow a Schmidt law,
with exponent n=1.4, and a density threshold. It is
calibrated for a gas consumption time-scale of 5 Gyr.
We do not assume instantaneous recycling, but
take into account a continuous mass loss, across Gyrs \citep{2001A&A...376...85J}.
The mass loss by stars is distributed through gas on
neighboring particles, with a velocity dispersion of 10km/s,
to schematize the feedback energy.

The RCs are fitted to the observedtypical curves for these Hubble types. For these giant spiral galaxies, the total mass inside 35kpc
is 2.4 10$^{11}$ M$_\odot$, with 75\% n DM and 25\% baryons,
for the standard model, and almost equality between baryons and DM
for the disk-dominated (dmx) model. According to the Hubble types, Sa, Sb, Sd, 
the bulge to disk ratio is varied.

Similar simulations with TREE-SPH have also been run,
with the same initial conditions (c.f. \citealt{2007A&A...468...61D}).
In this code, gravitational forces are calculated using a 
hierarchical tree method \citep{1986Natur.324..446B} and gas evolution is followed by
means of smoothed particle hydrodynamics (SPH, \citealt{1977AJ.....82.1013L};
\citealt{1982JCoPh..46..429G}). Gravitational forces are calculated
using a tolerance parameter $\theta =0.7$
and include terms up to the
quadrupole order in the multipole expansion. A Plummer
potential is used to soften gravitational forces, with same 
softening lengths for all particle types, of 250pc.
The code and the various validation tests
were described in \cite{2002A&A...388..826S}.
In the present study, we use the isothermal gas phase 
(gas temperature of 10$^4$K), adopted for the GalMer runs
\citep{2010A&A...518A..61C}.
To capture shocks a conventional form of the artificial 
viscosity is used, with parameters
$\alpha=0.5$ and $\beta=1.0$ \citep{1989ApJS...70..419H}. 
 Because of the short cooling time
of disc gas, fluctuations in the gas temperature are quickly 
radiated away, so that simulations employing an isothermal 
equation of state are justified.

Star Formation and continuous stellar mass loss,
together with supernovae feedback are taken into account.
We parametrized the star formation efficiency for a SPH particle as
$\frac{dM_{gas}}{dt M_{gas}} = C \rho^{1/2}$, with the constant
C chosen such that the unperturbed
giant disc galaxies form stars with a gas depletion time-scale of
2 Gyr. This is equivalent to a Kennicutt-Schmidt law of exponent 1.5.

To avoid to form too small stellar particles, we adopt the
method of hybrid particles, where progressively the gas particles
contain a stellar fraction for a certain time-scale, until the 
remaining gas fraction falls below 5\%. Then the particle is transferred
to the stellar component, and the gas 
is distributed over neighboring hybrid particles. 

The stellar feedback is implemented both with 
stellar mass loss, and through supernovae. It is assumed that
a percentage of 1\% of the young stellar population is becoming a supernova,
i.e. all stars more massive than 8 M$_\odot$, and the assumption of
instantaneous recycling. The released mass
also enriches the metallicity of the surrounding gas. 

The stellar mass-loss at each time step, is accounted for by 
the method described by \cite{2001A&A...376...85J}, and allows
 a more progressive metal enrichment of the interstellar
medium.
\section{Results from the SPH simulation}

\begin{figure*}
\centerline{
\includegraphics[angle=0,width=22cm]{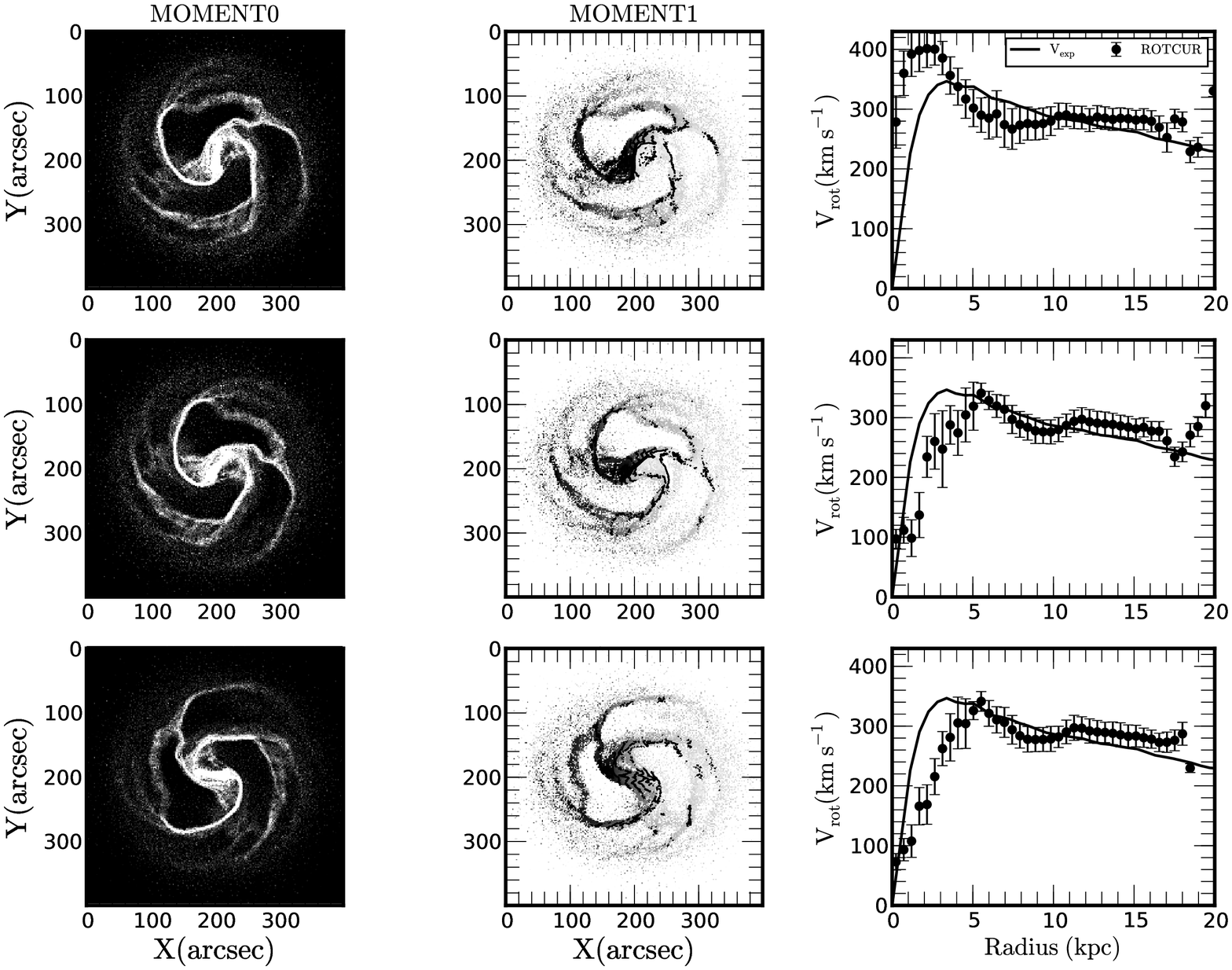}}
\caption{Three different bar positions for the model gSa (SPH simulation, T= 200 Myrs). Top panel: the bar is perpendicular to the major axis, middle panel: intermediate position and bottom panel: parallel to the major axis. 
The first column is the moment0 map, second column the moment 1 map superposed with the iso-velocity contours and the third 
column the comparison between the expected RCs shown as a continuous line and the measured rotational velocities derived using {\sc rotcur} (black points). } 
\label{sph1}
\end{figure*}

\begin{figure*}
\centerline{
\includegraphics[angle=0,width=22cm]{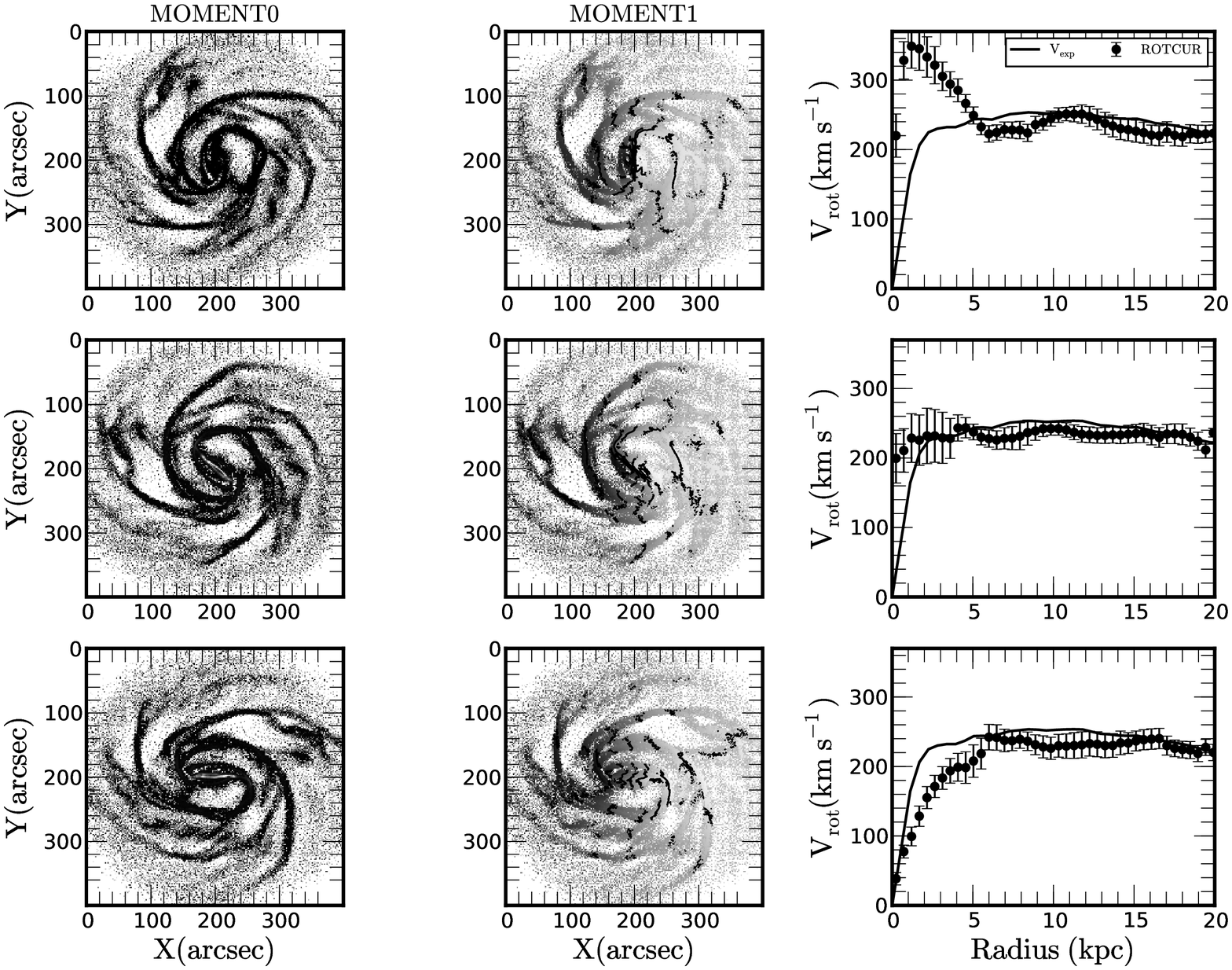}}
\caption{Three different bar positions for the model  gSb ( SPH simulation, T = 400 Myrs see figure  \ref{sph1} for details).}
\label{sph2}
\end{figure*}

The dynamics of the gas in disk galaxies are very complex systems and difficult to reproduce in numerical simulations. Therefore, to test the influence of our assumptions on the gas treatment on our results,
we have considered the same simulations, with the same initial conditions,
but run with the TREE-SPH code, used in the GalMer data-base (cf \citealt{2010A&A...518A..61C}). The main features of the SPH-tree code for this particular simulation and also  our treatment of star formation and feedback, are described in the appendix  \ref{apa}. The gas, which was considered essentially as an ensemble of dense clouds, dissipating their kinetic energy in collisions,  is now considered as a fluid with pressure forces and shocks. The simulations have twice more particles, and the softening length is twice smaller, so that the spatial resolution of the hydrodynamic is larger. The velocity maps and other results from the GalMer project is publicly available and can be downloaded from their website (\url{http://galmer.obspm.fr/}).
Figures \ref{sph1} and \ref{sph2} display the results with the TREE-SPH code for the gSa and gSb galaxies, corresponding to figure \ref{stck1} and \ref{stck2} with the FFT-sticky particles code. 
The gas behaviour is different, as expected, but the global feature
due to the bar, the perturbations of the RC, are similar.
For the orientation where the bar is parallel to the line of sight, the discrepancy
between the true circular velocity and the "observed" rotational velocity is even larger, since the spatial resolution is higher. In the FFT-sticky particles simulations, the high velocity gradient was a bit smoothed out. This is less visible, when the bar is perpendicular to the line of sight, which is a configuration reducing the "observed" velocity gradient. Therefore, this shows that using different simulation codes does not affect our results. 


\bibliographystyle{aa}

\bsp

\label{lastpage}

\end{document}